\documentclass[preprint,authoryear,12pt]{elsarticle}

\usepackage{graphicx}
\usepackage{subfig}
\usepackage{caption}
\usepackage{enumitem}
\usepackage{longtable}
\usepackage{amssymb}
\usepackage{amsmath}
\usepackage{physics}
\usepackage{mathtools}
\usepackage[breaklinks=true,colorlinks=true]{hyperref}
\usepackage[authoryear]{natbib}
\usepackage{pgf}
\usepackage{tikz}
\usetikzlibrary{positioning,fit,backgrounds}

\newcommand{\codingC}{\textsc{C}}
\newcommand{\wrt}[1]{\mathrm{d}{#1}}
\newcommand{\matlab}{\textsc{Matlab}}

\newcommand{\ode}[1]{\textsc{ode#1}}

\newcommand{\numpy}{\textsc{NumPy}}

\definecolor{red}{HTML}{FF1E00}
\definecolor{orange}{HTML}{FF7F00}
\definecolor{green}{HTML}{00BF32}
\definecolor{blue}{HTML}{04819E}
\definecolor{lightblue}{HTML}{7AC5D6}
\definecolor{lightorange}{HTML}{FFC58C}

\journal{Advances in Space Research}

\begin{document}

\begin{frontmatter}
	\title{Extension of the King-Hele orbit contraction method for accurate, semi-analytical propagation of non-circular orbits}

\author[add1]{Stefan Frey\corref{cor1}}
\ead{stefan.frey@polimi.it}
\author[add1]{Camilla Colombo}
\ead{camilla.colombo@polimi.it}
\author[add2]{Stijn Lemmens}
\ead{stijn.lemmens@esa.int}

\address[add1]{Department of Aerospace Science and Technology, Politecnico di Milano, Via La Masa, 34, 20156 Milan, Italy}
\address[add2]{ESA/ESOC Space Debris Office, Robert-Bosch-Str. 5, 64293 Darmstadt, Germany}
\cortext[cor1]{Corresponding author}
	\begin{abstract}
Numerical integration of orbit trajectories for a large number of initial conditions and for long time spans is computationally expensive. Semi-analytical methods were developed to reduce the computational burden. An elegant and widely used method of semi-analytically integrating trajectories of objects subject to atmospheric drag was proposed by King-Hele (KH).
However, the analytical KH contraction method relies on the assumption that the atmosphere density decays strictly exponentially with altitude.
If the actual density profile does not satisfy the assumption of a fixed scale height, as is the case for Earth's atmosphere, the KH method introduces potentially large errors for non-circular orbit configurations.

In this work, the KH method is extended to account for such errors by using a newly introduced atmosphere model derivative.
By superimposing exponentially decaying partial atmospheres, the superimposed KH method can be applied accurately while considering more complex density profiles.
The KH method is further refined by deriving higher order terms during the series expansion. A variable boundary condition to choose the appropriate eccentricity regime, based on the series truncation errors, is introduced.
The accuracy of the extended analytical contraction method is shown to be comparable to numerical Gauss-Legendre quadrature. Propagation using the proposed method compares well against non-averaged integration of the dynamics, while the computational load remains very low.

\end{abstract}

\end{frontmatter}

\parindent=0.5 cm

\section{Introduction}
Numerical integration of the full orbital dynamics, including short-periodic variations, can be demanding from a computational point of view. For this reason, Semi-Analytical (SA)\footnote{The abbreviations used herein are, in alphabetical order; CIRA: COSPAR International Reference Atmosphere, CNES: Centre National d'\'{E}tudes Spatiales, COSPAR: Committee on Space Research, DTM: Drag Temperature Model, GL: Gauss-Legendre, KH: King-Hele, NA: Non-Averaged, NRLMSISE: Naval Research Laboratory Mass Spectrometer, Incoherent Scatter Radar Extended, SA: Semi-Analytical, SI-KH: SuperImposed King-Hele} methods were developed to perform this task in a less demanding manner~\citep[e.g.][]{Liu1974}. Such methods remove the short-term periodic effects by averaging the variational equations, thereby reducing the stiffness of the problem. This is especially desired when orbits are to be propagated for many initial conditions and over long lifetimes, e.g. for estimating the future space debris environment.

The calculation of the orbit contraction -- i.e. the reduction in semi-major axis and eccentricity -- induced by atmospheric drag requires the integration of the atmosphere density along the orbit.
Half a century ago, King-Hele (KH) derived analytical approximations to these integrals~\citep{King1964}.
Depending on the eccentricity of the orbit, e.g. circular, near-circular, low eccentric and highly eccentric, different series expansions were derived.
Recommendations are given, found empirically, on when to use which formulation.
\citet{Vinh1979} improved the theory by removing the ambiguity arising from the regions of validity in eccentricity and by applying the more mathematically rigorous Poincar\'{e} method for integration.
The classical theory was adapted to non-singular elements, mitigating the problems that theories formulated in Keplerian elements have with vanishing eccentricities~\citep{Sharma1999, XavierJamesRaj2006}.

The advantage of these methods is that the averaged contraction can be computed analytically using only a single density evaluation at the perigee. However, the analytical methods assume exponential decay of the atmosphere density above the perigee height.
This fixed scale height assumption potentially introduces large errors, especially for highly eccentric orbits, if compared to propagation using quadrature.

Averaging methods based on quadrature solve the integral numerically.
No assumption on the shape of the density profile is required, however, the density needs to be evaluated at many nodes along the orbit, slowing down the integration of the trajectory.

This work proposes modelling the atmosphere density by superimposing exponential functions, each with a fixed scale height.
The KH formulation is then used for the calculation of the contraction of each individual component.
As the assumption of a fixed scale height is satisfied for each component, the resulting decay rate is estimated with great accuracy.
Finally, each individual contribution is summed up, resulting in the global contraction of the overall not strictly exponentially decaying atmosphere density.
This superimposed approach is not limited to the KH method and can also be applied to the other analytical methods described above.

The proposed method is applied during propagation of different initial conditions from circular to highly elliptical orbits and compared against propagations using numerical quadrature of the contraction as well as against Non-Averaged (NA) integration.
The smooth atmosphere derivative introduced here is independent of the underlying atmosphere model and can be extended to include time-variations, as shown here for the case of solar activity.
\section{Background on Atmospheric Models}
\label{sec:background_atmospheremodels}
The atmosphere models discussed here can be divided into reference models and the derivatives thereof.
The reference models commonly give the temperature, $T$\footnote{The nomenclature of all the variables can be found in \ref{sec:nomenclature}.}, and -- more importantly for calculating the drag force -- the density, $\rho$, of Earth's atmosphere as a function of the altitude, $h$, and other input parameters.
Examples are, in increasing degree of complexity, the COSPAR International Reference Atmosphere (CIRA), the Jacchia atmosphere~\citep{Jacc1977}, the Drag Temperature Model (DTM)~\citep{Bruinsma2015} and the Naval Research Laboratory Mass Spectrometer, Incoherent Scatter Radar Extended (NRLMSISE) model~\citep{Pico2002}, all of which are (semi-)empirical models.

Of these reference models, derivatives can be obtained through fitting for two purposes: appropriate simplification of the mathematical formulation can lead to significant speed increases for a density evaluation; and adequate reformulation of the model improves the accuracy of analytical SA contraction methods as will become apparent in Section~\ref{sec:kinghele}.

Sections~\ref{sec:jacchia} and \ref{sec:nonsmooth} briefly introduce the Jacchia-77 atmosphere model and a derived non-smooth exponential atmosphere model.

\subsection{Jacchia-77 Reference Atmosphere Model}
\label{sec:jacchia}
The Jacchia-77 reference atmosphere~\citep{Jacc1977} estimates the temperature and density profiles of the relevant atmospheric constituents as a function of the exospheric temperature, $T_\infty$.
The density profile, $\rho$${}_J$, is based on the barometric equation and an empirically derived temperature profile in order to comply with observations of satellite decay.
The static model is valid for altitudes $90 < h < 2500$~km and exospheric temperatures $500 < T_\infty < 2500$~K.

The computation of $\rho$${}_J$ cannot be performed analytically and requires numerical integration for each of the 4 constituents, nitrogen, oxygen, argon and helium, plus integration of atomic nitrogen and oxygen.
A fast, closed-form approximation is available~\citep{Lafontaine1983}, but it was not considered here, as its modelled atmosphere does not purely decay exponentially.

The scale height, $H$, is defined as
\begin{equation}
	\label{eq:scaleheight}
	H = -\frac{\rho}{\wrt{\rho}/\wrt{h}}
\end{equation}
and numerically approximated for the Jacchia model scale height,~$H$${}_J$, as
\begin{equation}
\label{eq:scaleheightjacchia}
	H_J(h) = -\frac{\rho(h) \Delta h}{\rho(h+\Delta h)-\rho(h)}		\qquad \Delta h=1 \text{~m}
\end{equation}

\begin{figure}
	\centering
	\includegraphics[width=\linewidth]{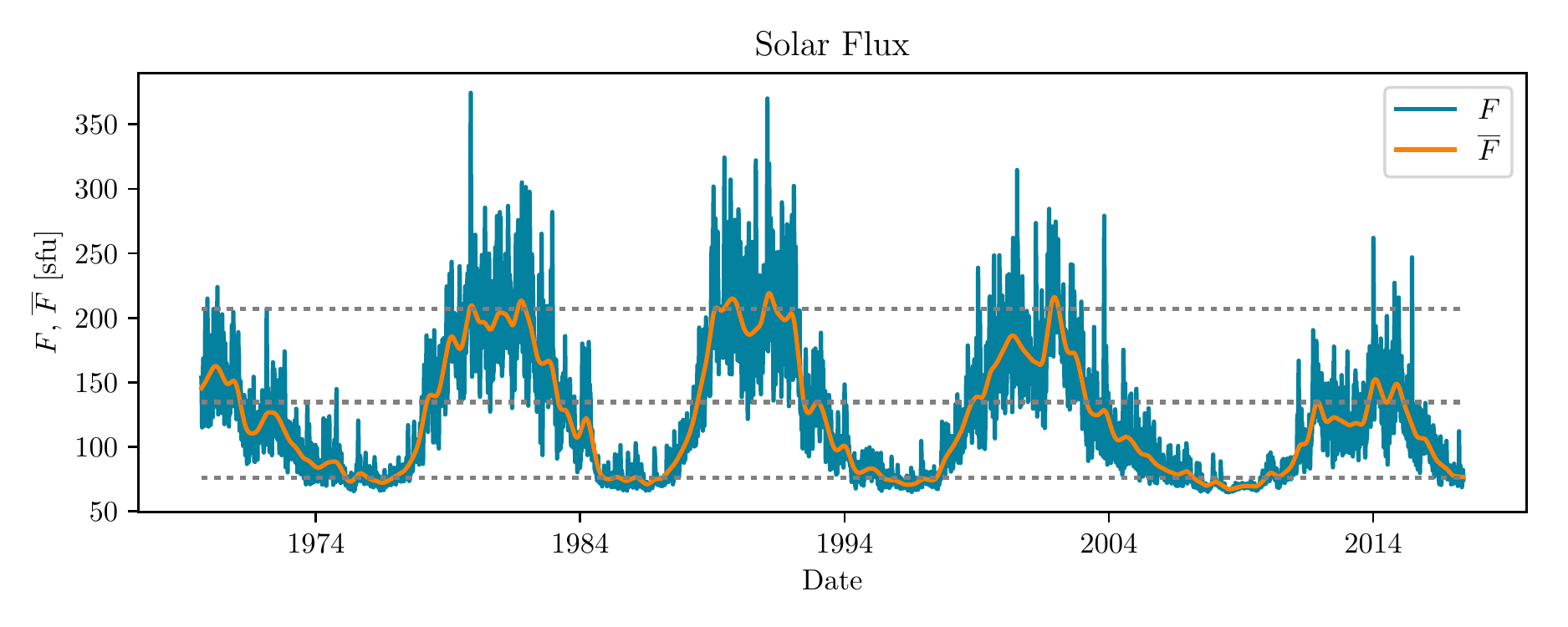}
	\caption{Daily $10.7$~cm solar flux, and a Gaussian mean with $\sigma=81$~days and a window of $w=486$~days, since beginning of 1970. The dashed lines correspond to $T_\infty=750$, $1000$ and $1250$~K, respectively, assuming $F=\overline{F}$.}
	\label{fig:solarflux}
\end{figure}

Several thermospheric variations can be taken into account, such as solar cycle, solar activity, seasonal or daily variations.
Generally, the objects of interest for SA propagation dwell on-orbit for several months to hundreds of years.
Thus, only the variation with the $11$-year solar cycle is of interest here.
The Jacchia reference uses the solar radio flux at $10.7$~cm, $F$, as an index for the solar activity~\citep[see Figure~\ref{fig:solarflux}, source for data:][]{Goddard}. From $F$, $T_\infty$ can be inferred as~\citep{Jacc1977}
\begin{equation}
	\label{eq:Tinf}
	T_\infty = 5.48 \overline{F}^\frac{4}{5} + 101.8 F^\frac{2}{5}
\end{equation}
where $\overline{F}$ is a smoothed $F$, commonly centred over an interval of several solar rotations. Jacchia recommended to use a smooth Gaussian mean based on weights which decay exponentially with time. Figure~\ref{fig:solarflux} shows the solar flux,~$F$, and the Gaussian mean with a standard deviation of $\sigma=3$~solar rotations, i.e. 81~days, considering a window of $\pm 3 \sigma$. More recent models such as NRLMSISE or DTM require $\overline{F}$ to be a moving mean of 3~solar rotations~\citep{ISO2013}.

\subsection{Non-Smooth Exponential Atmosphere Model}
\label{sec:nonsmooth}
One very simple representation of the atmosphere density is using a piece-wise exponentially decaying model, by dividing the altitude range into bins.
Each bin is defined by a lower altitude (base) and an upper altitude (base of the next bin), $h$$_i$ and $h$$_{i+1}$, respectively, the base density, $\hat{\rho}$$_i$, at $h$$_i$ and a scale height, $H$$_i$, chosen such that the density is continuous over the limits of each bin.
Then, within each altitude bin, the density, $\rho$${}_{NS}$, can be evaluated at each altitude $h$ as follows
\begin{equation}
	\rho_{NS}(h) = \hat{\rho}_i \exp{\frac{h-h_i}{H_i}} \qquad h_i < h < h_{i+1}
\end{equation}
Such a model can be derived from any atmospheric model.
Herein, the values given in~\citet[][Chapter 8.6]{Vall2013} -- fitting the CIRA-72 model at $T_\infty$$=1000$~K -- are used for a comparison of models.

A problem with the non-smooth atmosphere model is that it is non-physical, with discontinuities in $H$. 
At each change of altitude bin, $H$ jumps from $H$$_i$ to $H$$_{i+1}$. 
This non-smooth behaviour poses a problem to the (variable-step size) integrator, as the step size needs to be reduced to accurately describe the sudden change in contraction rate of the orbit. 
Thus, the number of function evaluations and the total time to propagate the orbit increases.
An example is given in Figure~\ref{fig:ContVsNumSteps}, comparing the number of steps required for propagation of an object subject to the non-smooth~$\rho$${}_{NS}$ to one using the smooth~$\rho$${}_J$ as a function of altitude. 
Evidently, each change of bin forces the integrator to reduce the step size.

The equally simple parametric model introduced in Section~\ref{sec:smoothAtmosphere} does not suffer from these discontinuities.
\section{Background on Semi-Analytical Orbit Contraction Methods}
\label{sec:background_semianalytic}
During SA propagation of an object trajectory subject to air-drag forces, the integrated change in the orbital element space, i.e. the contraction of the orbit, over a full revolution is of interest.
This requires the integration of the (weighted) density along the orbit, which can either be done numerically using quadrature, or analytically.

Many quadrature rules exist \citep[e.g. see][p.~885--895]{Abramowitz1964} and they are independent of the underlying function, making them versatile.
However, they require the evaluation of the density at multiple nodes along the orbit, increasing the computational load of the function evaluations during integration.

Analytical formulations, such as the one derived by D. King-Hele more than half a century ago~\citep{King1964} require the density to be evaluated only once per iteration in correspondence of the perigee altitude.
Other examples of analytical formulations are the ones derived by~\citet{Vinh1979}, \citet{Sharma1999} and~\citet{XavierJamesRaj2006}.
While offering improvements to the classical formulation of KH, such as being mathematically more rigorous and non-singular, they still suffer from the same assumption of a fixed scale height.
The method proposed in Section~\ref{sec:extension} addresses this problem for any of the analytical formulations.
For the sake of brevity, it is only applied to the KH method. 

Section~\ref{sec:dynsystem} introduces the system dynamics used throughout this work and discusses its averaging. Sections~\ref{sec:numapprox} and \ref{sec:kinghele} introduce two averaging methods; the numerical Gauss-Legendre (GL) quadrature and the analytical KH method.

\subsection{Dynamical System and Averaging}
\label{sec:dynsystem}
The main focus of this work is on correcting the errors arising from the fixed scale height assumption.
Important effects of an oblate Earth, such as a non-spherical atmosphere or gravitational coupling~\citep[e.g. see~][]{Brower1961}, are not considered here.
The superimposed approach does not replace the averaging method, rather it transforms one of the inputs, i.e. the atmosphere density, to fit its assumptions.
Hence, it is also applicable to more elaborate theories.

The dynamical system used here is based on Lagrange's planetary equations, given in Keplerian elements, stating the changes in the elements as a function of the applied forces from any small perturbations~\citep[see][for more information]{King1964}.
Only the tangential force induced by the aerodynamic drag is considered, i.e.
\begin{equation}
\label{eq:dragforce}
f_T = -\frac{1}{2} \rho v^2 \delta
\end{equation}
with the density, $\rho$, the inertial velocity, $v$, and the effective area-to-mass ratio (i.e. the inverse of the ballistic coefficient), $\delta$, defined as $\delta=c_D A/m$, where $c_D$ is the drag coefficient, $A$ is the surface normal to $v$, and $m$ is the mass.
Atmospheric rotation is ignored here, but could be taken into account by multiplying the right hand side of Equation~\ref{eq:dragforce} with the appropriate factor.

The variations of the semi-major axis, $a$, the eccentricity, $e$, and the eccentric anomaly, $E$, with respect to time, $t$, are
\begin{subequations}
	\label{eq:modLagrangePlanetary}
	\begin{align}
		\dv{a}{t} &= -\frac{a^2 \rho \delta v^3}{\mu} \\
		\dv{e}{t} &= \frac{a \rho \delta v}{r} (1-e^2) \cos{E} \\
		\dv{E}{t} &= \frac{1}{r} \left( \frac{\mu}{a} \right)^{\frac{1}{2}}
	\end{align}
\end{subequations}
with Earth's gravitational parameter, $\mu$, the radius, $r$, and $v$ given as
\begin{subequations}
	\begin{align}
	r &= a (1-e \cos{E}) \\
	v &= \sqrt{2 \frac{\mu}{r}-\frac{\mu}{a}}
	\end{align}
\end{subequations}

In order to reduce the stiffness of the problem, Equation~\ref{eq:modLagrangePlanetary} is averaged over a full orbit revolution, under the assumption that $a$ and $e$ remain constant. The resulting contractions, $\Delta a$ and $\Delta e$, for $a$ and $e$ respectively are
\begin{subequations}
	\label{eq:averagedLagrangePlanetary}
	\begin{align}
	\Delta a &= -a^2 \delta \int_{0}^{2 \pi} \rho(h) \frac{(1+e \cos{E})^{\frac{3}{2}}}{(1-e \cos{E})^{\frac{1}{2}}} \wrt{E} \\
	\Delta e &= -a \delta \int_{0}^{2 \pi} \rho(h) \left( \frac{1+e \cos{E}}{1-e \cos{E}} \right)^\frac{1}{2} \cos{E} (1-e^2) \wrt{E}
	\end{align}
\end{subequations}
with the altitude, $h=r-R$, given the mean Earth radius, $R$.

For SA propagation of the orbit, the derivatives of the variables with respect to time are approximated by the change over one revolution divided by the time required to cover the revolution
\begin{equation}
	\label{eq:rateofchange}
	F_x = \dv{x}{t} \approx \frac{\Delta x}{P} \qquad x \in [a, e]
\end{equation}
with the orbit period, $P$, defined as
\begin{equation}
	P = 2 \pi \sqrt{ \frac{a^3}{\mu} }
\end{equation}

\subsection{Numerical Approximation}
\label{sec:numapprox}
The integrals in Equation~\ref{eq:averagedLagrangePlanetary} can be approximated numerically using quadrature, e.g. GL quadrature~\citep[][p.~887]{Abramowitz1964}
\begin{equation}
\int_0^{2 \pi} f(E) dE \approx \pi \sum_{i} w_i f(E_i), \quad E_i =  (x_i+1) \pi
\end{equation}
where the node $x_i$ is the $i^{th}$ root of the Legendre Polynomial $P_n(x)$. The weights $w_i$ are given as
\begin{equation}
w_i = \frac{2}{(1-x_i^2)[P'_n(x_i)]^2}
\end{equation}
and $P'_n$ is the derivative of $P_n(x)$ with respect to $x$.
The nodes and weights remain constant during the propagation, so they are calculated (or read from a table) only once upon initialisation.
Routines to calculate ($x_i$, $w_i$) are available for various scientific programming tools, such as \matlab{}~\citep{MATLAB} and \numpy~\citep{NumPy}.

Advantages of a numerical approximation of the integrals in Equation~\ref{eq:averagedLagrangePlanetary} is that it can be found for any atmospheric model and that no series expansions are required.
Disadvantages are the need of multiple density evaluations and the loss of an analytic formulation. E.g. the Jacobian cannot be inferred analytically, but requires another quadrature.

\subsection{Classical King-Hele Approximation}
\label{sec:kinghele}
Here, only a brief summary of the formulation is given.
The treatment of the full theory behind the KH formulation can be found in~\citet{King1964}.
The integrals in Equation~\ref{eq:averagedLagrangePlanetary} can be approximated analytically by expanding the integrands as a power series in $e$ for low eccentric orbits, and in the inverted auxiliary variable, $z$
\begin{equation}
	\label{eq:auxilaryvariable}
	\frac{1}{z} = \frac{H}{a e}
\end{equation}
for highly eccentric orbits, and cutting off at the appropriate degree.

With the assumption that the density, $\rho$, decreases strictly exponentially with altitude, i.e. with a fixed $H$, each expanded integrand can be represented by the modified Bessel function of the first kind, $I_n$, which for $n \in \mathbb{N}_0$ is given as~\citep[][p.~376]{Abramowitz1964}
\begin{equation}
	\label{eq:besselfunction}
	I_n(x) = \frac{1}{\pi} \int_{0}^{\pi} \exp{(x \cos \theta) \cos{(n \theta)}} d\theta
\end{equation}
In~\ref{sec:KH}, the resulting equations are given up to $5$\textsuperscript{th} order, higher than the $2$\textsuperscript{nd} order given originally by~KH.
\begin{figure}
	\centering
	\subfloat[Drag forces are greatly underestimated using the KH formulation, resulting in slower orbital decay.]{
		\includegraphics[width=.47\linewidth]{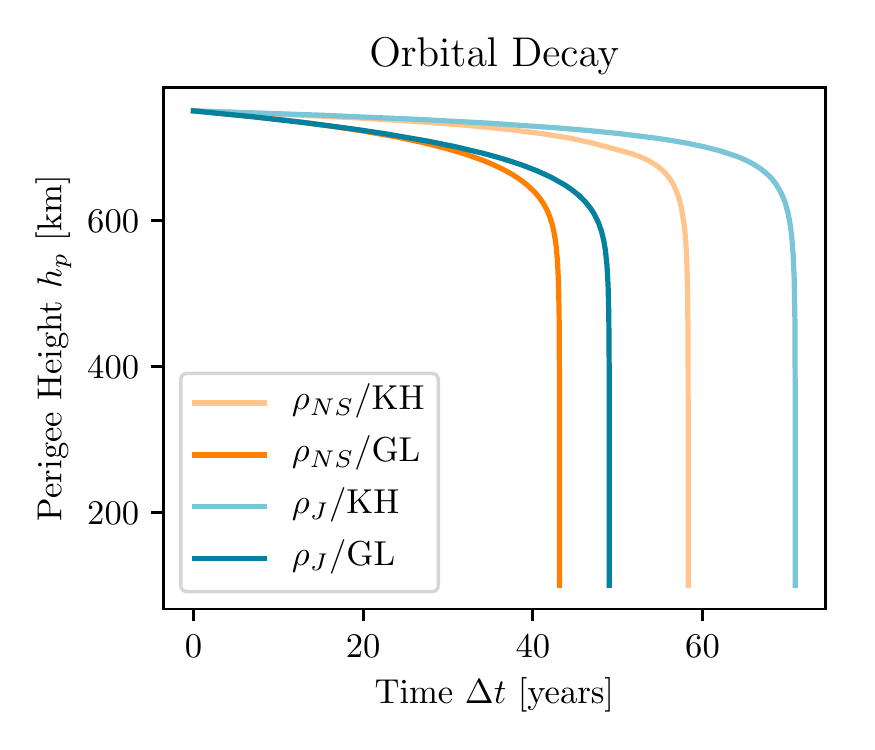}
		\label{fig:ContVsNumTrajectories}}
	\hfill
	\subfloat[The non-smooth density profile of $\rho$${}_{NS}$ forces the integrator to increase the number of steps, $N_s$.]{
		\includegraphics[width=.47\linewidth]{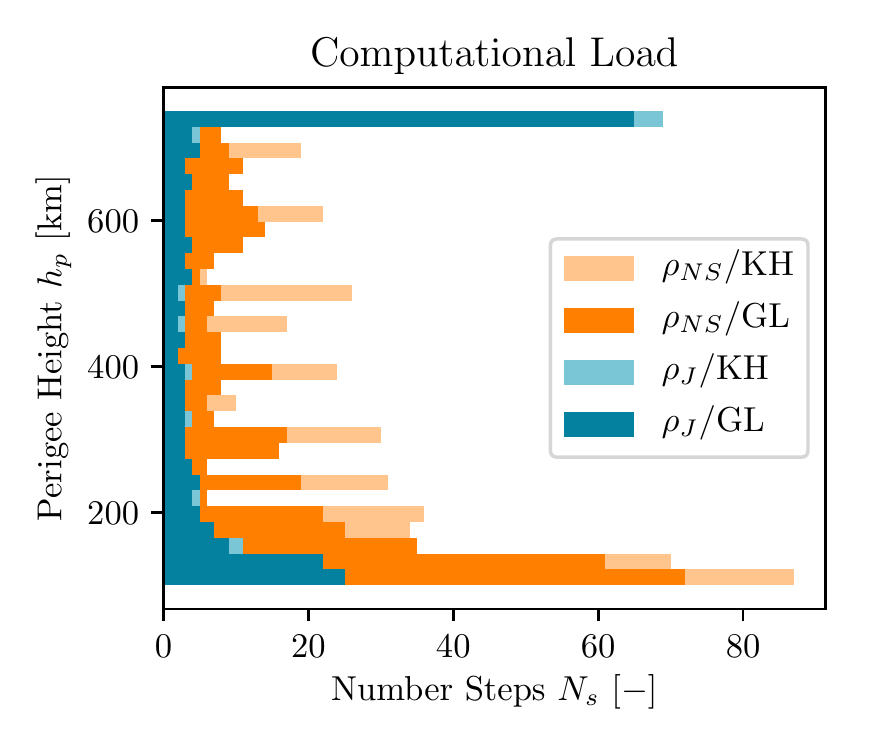}
		\label{fig:ContVsNumSteps}}
	\caption{Trajectories propagated for two different atmosphere models, $\rho$${}_{NS}$ (orange) and $\rho$${}_J$ (blue), and two different contraction methods, KH (light) and GL (dark). The initial state is $h_p \times h_a=750 \times 2000$~km.}
	\label{fig:ContVsNum}
\end{figure}

The KH formulation is fast as it can be evaluated analytically and requires only a single density evaluation for each computation of the contraction.
The main problem with the fixed $H$ assumption is the underestimation of $\rho$ at altitudes above the perigee altitude, $h_p$, which for eccentric orbits can induce large errors.
Figure~\ref{fig:ContVsNumTrajectories} shows the trajectories of an object in an initially eccentric orbit with perigee and apogee height of $h_p \times h_a=750\times2000$~km.
They were propagated with two different atmosphere models, $\rho$${}_{NS}$ and $\rho$${}_J$, and using two different contraction methods, GL quadrature and the KH formulation.
For both atmosphere models, the KH method overestimates the density decay above perigee along the orbit, leading to an overestimation of the lifetime of up to $40 \%$, compared to the propagation with the GL method.
This is true -- albeit sometimes less pronounced -- for any object in a non-circular orbit subject to a non-strictly exponentially decaying atmosphere.

It has to be noted here that KH was aware of this problem and suggested a way to calculate the contraction of an orbit with a varying scale height~\citep[see][Chapter 6]{King1964}. 
To keep the equations analytically integrable, he approximates the varying $H$ linearly, with a constant slope parameter.
Linear approximation of the true $H$ is valid only locally. 
For low eccentric orbit configurations this might be sufficient, but high eccentricities will re-introduce the errors.
Using a constant slope parameter will thus lead to a new over- or underestimation of the drag depending on $e$. 

Another issue of the KH formulation is that it relies on series expansion.
As the eccentricity grows, the formulation to calculate the contraction needs to switch from low to high eccentric orbits.
This introduces discontinuities, at a classically fixed boundary eccentricty, $e_b$.
\section{Proposed new model for the semi-analytical computation of the orbit contraction due to atmospheric drag}
\label{sec:extension}
The proposed method of taking into account atmospheric drag for SA integration of trajectories consists of two parts: an atmosphere model based on constant scale heights, introduced in Section~\ref{sec:smoothAtmosphere}; and the extension of the KH formulation to reduce the errors induced by an atmosphere which in its sum does not decay exponentially, described in Section~\ref{sec:SIKH}.
\begin{table}
	\centering
	\begin{tikzpicture}[node distance=1mm]

\tikzstyle{title}=[minimum height=.65cm, minimum width=6.8cm]
\tikzstyle{subtitle}=[minimum height=.65cm, minimum width=3.2cm]
\tikzstyle{element1}=[minimum height=.61cm, minimum width=3.2cm, align=center,fill=lightblue]
\tikzstyle{element2}=[element1,fill=lightorange]

\node[title, draw=lightblue, fill=lightblue] (atmos) at (0,0) {Atmosphere models: $\rho$};

\node[subtitle, below = of atmos.south,anchor=north east, xshift=-3pt, yshift=-1mm] (references) {References};
\node[element1, below = of references] (cira) {CIRA};
\node[element1, below = of cira] (jacchia) {Jacchia};
\node[element1, below = of jacchia] (nrlmsise) {NRLMSISE};
\node[element1, below = of nrlmsise] (dtm) {DTM};
\node[draw=lightblue, inner sep=2pt, fit={(references) (cira) (jacchia) (nrlmsise) (dtm)}] (refbox) {};

\node[subtitle, below = of atmos.south,anchor=north west, xshift=3pt, yshift=-1mm] (derivatives) {Derivatives};
\node[element1, below = of derivatives] (nonsmooth) {Non-smooth \\ exponential};
\node[element1, below = of nonsmooth] (smooth) {Smooth \\ exponential};
\node[draw=lightblue, inner sep=2pt, fit={(derivatives) (nonsmooth) (smooth)}] (derbox) {};

\node[title, draw=lightorange, fill=lightorange, right = of atmos] (contr) {SA contraction methods: $\Delta a$, $\Delta e$};

\node[subtitle, below = of contr.south,anchor=north east, xshift=-3pt, yshift=-1mm] (analytical) {Analytical};
\node[element2, below = of analytical] (kinghele) {King-Hele (KH)};
\node[element2, below = of kinghele] (vinh) {Vinh et al.};
\node[element2, below = of vinh] (sharma) {Sharma};
\node[element2, below = of sharma] (superimposed) {Superimposed \\ KH (SI-KH)};
\node[draw=lightorange, inner sep=2pt, fit={(analytical) (kinghele) (vinh) (sharma) (superimposed)}] (anabox) {};

\node[subtitle, below = of contr.south,anchor=north west, xshift=3pt, yshift=-1mm] (numerical) {Numerical};
\node[element2, below = of numerical] (simpson) {Simpson's rule};
\node[element2, below = of simpson] (gauss) {Gauss- \\ Legendre (GL)};
\node[draw=lightorange, inner sep=2pt, fit={(numerical) (simpson) (gauss)}] (numbox) {};

\end{tikzpicture}\unskip
	\caption{Non-exhaustive list of existing and newly proposed atmospheric models and contraction methods.}
	\label{tab:overview}
\end{table}

Table~\ref{tab:overview} shows an overview of how the proposed extension fits into the existing scheme of atmosphere models and SA orbit contraction methods.
As mentioned earlier, the technique presented here is not limited to the KH method, but could be applied to any averaging method which is based on the fixed scale height assumption.

\subsection{Smooth Exponential Atmosphere Model}
\label{sec:smoothAtmosphere}
The smooth atmosphere model proposed here does not in any way attempt to replace existing atmosphere density models.
Instead, it is a derivation of those models.
Nor is the idea of modelling the atmosphere as a sum of exponentials new: the Jacchia-77 reference model reduces -- for each atmospheric constituent -- to such a mathematical formulation if the vertical flux terms are neglected~\citep{Bass1980}.
The novelty of this work is the combination of the atmosphere model with the extended, superimposed KH formulation. Sections~\ref{sec:staticAtmosphere} and~\ref{sec:variableAtmosphere} introduce the static and variable atmosphere model, respectively.

\subsubsection{Static Model}
\label{sec:staticAtmosphere}
The smooth exponential atmosphere model, $\rho$${}_S$, is modelled by superimposing exponentials functions as
\begin{equation}
	\label{eq:smoothAtmosphere}
	\rho_S(h) = \sum_{p=1}^{n_p} \rho_p(h) = \sum_{p=1}^{n_p} \hat{\rho}_p e^{-h/H_p}
\end{equation}
where the number of partial atmospheres, $n_p$, the partial base densities, $\hat{\rho}$${}_p$, and the partial scale heights, $H$${}_p$, are fitting parameters.
Note that the subscript $p$ does not stand for altitude bins, but for one of the partial atmospheres, each of which is valid for the whole altitude range.
While it potentially could stand for a single atmosphere constituent, it is not restricted as such.
The superimposed scale height, $H$${}_S$, is 
\begin{equation}
	H_S(h) = -\frac{\rho_S(h)}{\wrt{\rho_S}/\wrt{h}} = \frac{\sum_{p=1}^{n_p} \rho_p(h)}{\sum_{p=1}^{n_p} \rho_p(h)/H_p}
\end{equation}
The derivative of $H$${}_S$ with respect to $h$ is monotonically increasing, as $H$${}_p$ is enforced to be larger than 0 for all $p$.
Hence, the smooth atmosphere model can only be fitted to atmosphere models in altitude ranges where $\frac{\wrt{H}}{\wrt{h}} > 0$. 
Above $h=100$~km, this is the case for $\rho$${}_J$ for a wide range of $T_\infty$.
Even if the underlying model shows slightly negative $H$ at the lower boundary $h_0$, a partial atmosphere with a small positive $H$${}_p$ can still be fitted accurately.

\begin{figure}
	\centering
	\includegraphics[width=.5\linewidth]{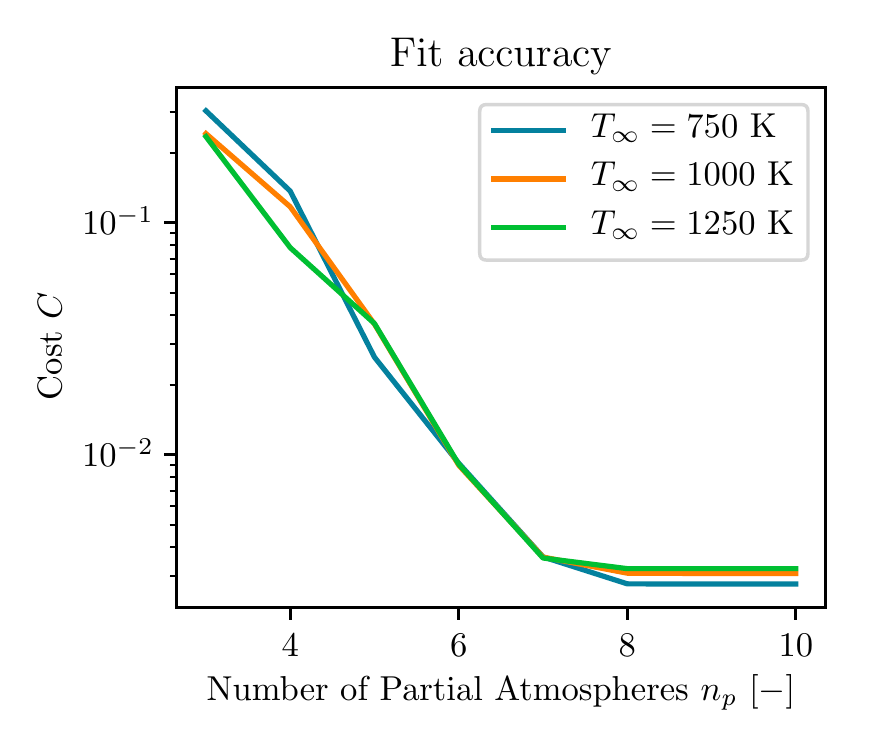}
	\caption{Cost function depending on number of partial atmospheres.}
	\label{fig:cost}
\end{figure}

\begin{figure}
	\centering
	\subfloat[Density profiles and density ratio.]{
		\includegraphics[width=.65\linewidth]{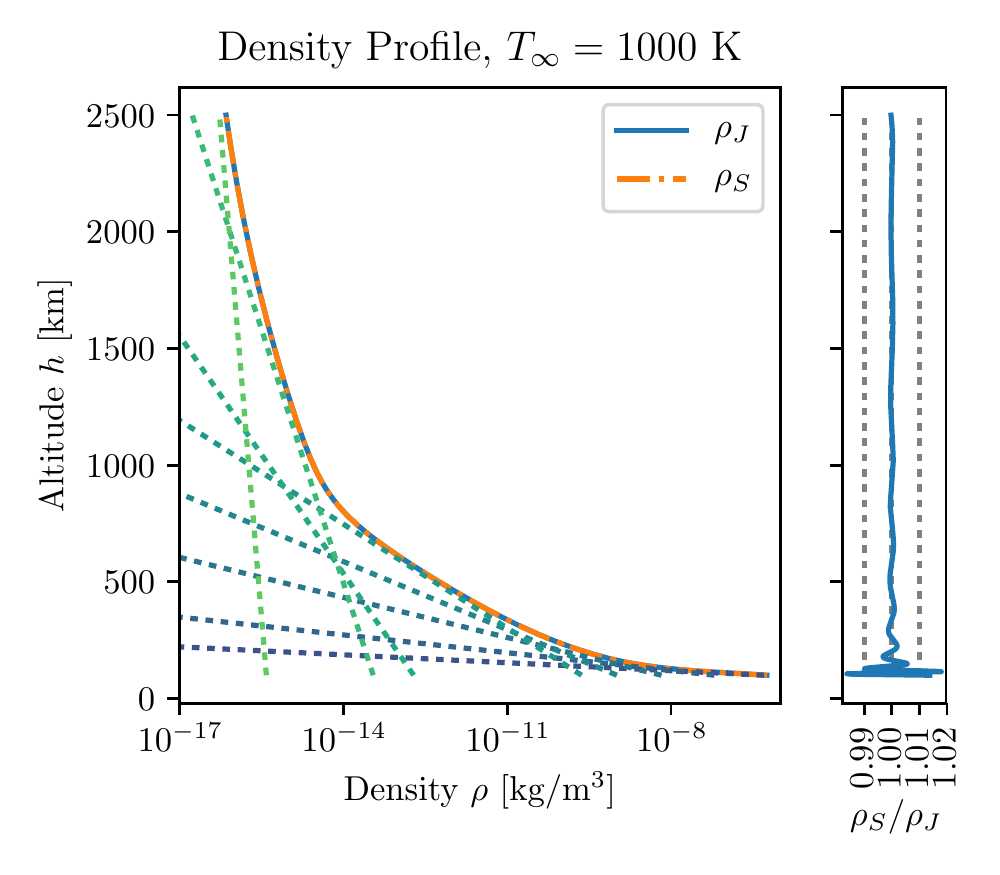}
		\label{fig:Fit1000Density}}

	\subfloat[Scale height profiles and scale height ratio. Note that the partial scale heights (dotted) are constant in $h$, while the superimposed scale height is not.]{
		\includegraphics[width=.65\linewidth]{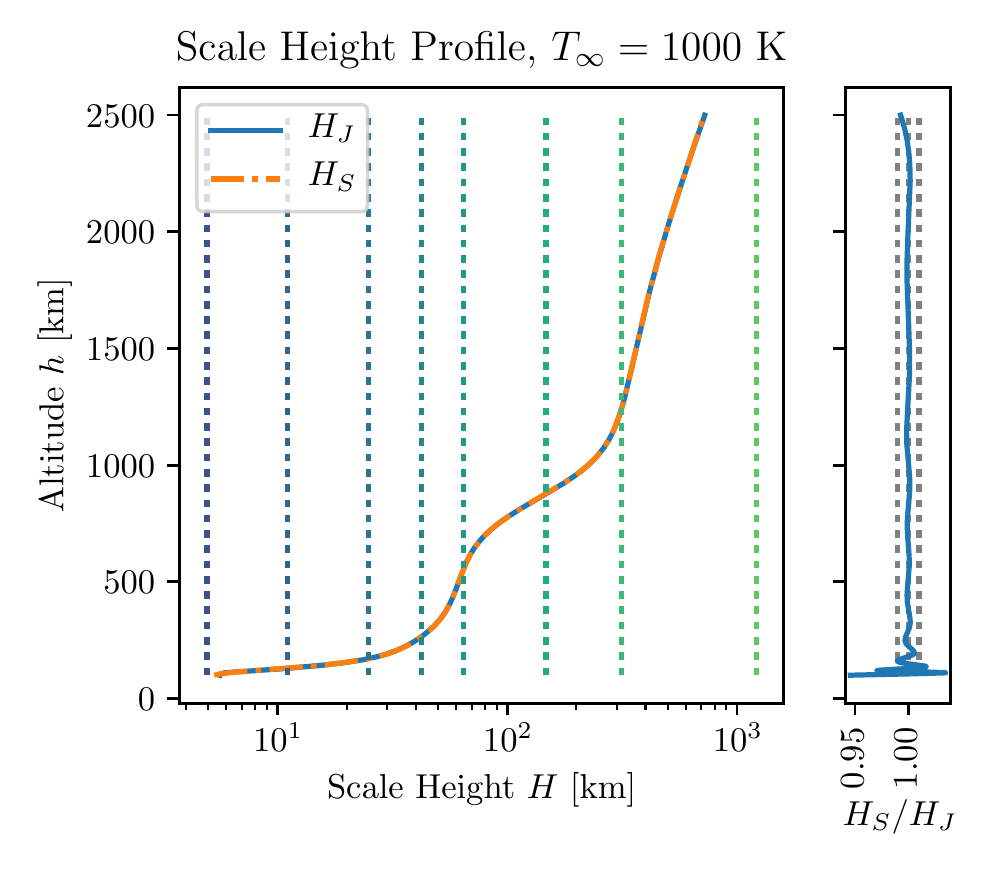}
		\label{fig:Fit1000ScaleHeight}}
	\caption{Fit of $\rho$${}_S$ to $\rho$${}_J$ for $T_\infty=1000$~K. Additionally, the different contributions of each partial atmosphere are shown (dotted) from $p=1$ (dark) to $p=8$ (light).}
	\label{fig:Fit1000}
\end{figure}

To find the parameters, $H$${}_p$ and $\hat{\rho}$${}_p$, the model in Equation~\ref{eq:smoothAtmosphere} is fitted to $\rho$${}_J$ for three different $T_\infty$:
in accordance to a low solar activity, $T_\infty=750$~K; mean solar activity, $T_\infty=1000$~K; and high solar activity, $T_\infty=1250$~K (see Figure~\ref{fig:solarflux}).
The fit is performed in the logarithmic space as not to neglect lower densities at higher altitudes, using least squares minimisation at heights between $h_0=100$~km and the upper boundary, $h_1=2500$~km.
To put more weights on the edges of the fit interval, the densities are evaluated at $N=100$ heights, $h_i$, distributed as Chebyshev nodes~\citep[][p.~889]{Abramowitz1964}
\begin{equation}
	\label{eq:Chebyshev}
	h_i = \frac{h_0+h_1}{2}+\frac{h_1-h_0}{2} \cos \left( \frac{2i-1}{2N} \pi \right) \qquad i=1, \dots, N
\end{equation}
The number of partial atmospheres, $n_p$, is chosen to be $8$, as the cost function
\begin{equation}
	C = \sqrt{\frac{1}{N}\sum_{i=1}^N \ln \left( \frac{\rho_S(h_i)}{\rho_J(h_i)} \right)^2}
\end{equation}
which is the root mean square of the logarithmic density fit residuals, stops improving (see Figure~\ref{fig:cost}). 
For $T_\infty \in [750, 1000, 1250]$~K, the relative error, $\eta_\rho$, calculated as
\begin{equation}
	\eta_\rho(h)=\frac{|\rho_S(h)-\rho_J(h)|}{\rho_J(h)}
\end{equation}
always remains below $0.1\%$ and $1\%$ for all $h > 308$~km and $h > 130$~km, respectively, and the maximum relative error, $\eta_\rho$$_{, max}$, does not exceed $2\%$, as can be seen in~Table~\ref{tab:fitErrors}.
Hence, the density fit accurately represents the underlying model.
The model parameters can be found in Table~\ref{tab:FitParameters}. 
Figure~\ref{fig:Fit1000} shows a comparison between the underlying and fitted model, for $T_\infty$$=1000$~K.

\begin{table}
	\centering
	\caption{Relative density fitting errors $\forall h \in [100, 2500]$~km.}
	\begin{tabular}{ l c c c } 
 \hline
 $\eta_\rho$			& $T_\infty=750$~K		& $T_\infty=1000$~K		& $T_\infty=1250$~K \\
 \hline
 $<0.1\%$			& $\forall h>239$~km	& $\forall h>308$~km	& $\forall h>306$~km \\
 $<0.5\%$			& $\forall h>134$~km	& $\forall h>153$~km	& $\forall h>154$~km \\
 $<1\%$				& $\forall h>119$~km	& $\forall h>119$~km	& $\forall h>130$~km \\
 $\eta_{\rho, max}$	& $1.6\%$ ($h=115$~km)	& $1.8\%$ ($h=115$~km)	& $1.9\%$ ($h=115$~km) \\
 \hline
\end{tabular}
	\label{tab:fitErrors}
\end{table}

\begin{table}
	\centering
	\caption{Smooth atmosphere model parameters resulting from a fit to the Jacchia-77 model, valid for altitudes $h \in [100, 2500]$~km.}
	\begin{tabular}{ l | c c | c c | c c } 
 \hline
 		& \multicolumn{2}{c | }{$T_\infty=750$~K} 	& \multicolumn{2}{c | }{$T_\infty=1000$~K}		& \multicolumn{2}{c}{$T_\infty=1250$~K} \\
 $p$	& $H_p$		& $\hat{\rho}_p$& $H_p$ 	& $\hat{\rho}_p$& $H_p$		& $\hat{\rho}_p$ \\
 		& [km]		& [kg/m$^3$]	& [km]		& [kg/m$^3$]	& [km]		& [kg/m$^3$] \\
 \hline
 $1$	& $4.9948$	& $2.4955\text{e}+02$	& $4.9363$	& $3.1632\text{e}+02$	& $4.9027$	& $3.6396\text{e}+02$ \\
 $2$	& $10.471$	& $8.4647\text{e}-04$	& $11.046$	& $5.2697\text{e}-04$	& $11.437$	& $3.8184\text{e}-04$ \\
 $3$	& $21.613$	& $9.1882\text{e}-07$	& $24.850$	& $3.7354\text{e}-07$	& $25.567$	& $2.8928\text{e}-07$ \\
 $4$	& $37.805$	& $1.2530\text{e}-08$	& $46.462$	& $1.0839\text{e}-08$	& $44.916$	& $1.2459\text{e}-08$ \\
 $5$	& $49.967$	& $1.3746\text{e}-09$	& $64.435$	& $1.0880\text{e}-09$	& $76.080$	& $9.2530\text{e}-10$ \\
 $6$	& $174.23$	& $1.5930\text{e}-13$	& $147.46$	& $3.8122\text{e}-13$	& $111.09$	& $1.6667\text{e}-11$ \\
 $7$	& $315.15$	& $1.1290\text{e}-14$	& $314.53$	& $4.8431\text{e}-14$	& $354.23$	& $5.9225\text{e}-14$ \\
 $8$	& $1318.1$	& $3.8065\text{e}-16$	& $1214.6$	& $4.2334\text{e}-16$	& $892.19$	& $1.7378\text{e}-15$ \\
 \hline
\end{tabular}
	\label{tab:FitParameters}
\end{table}

A speed test for $2401$ density and scale height evaluations over the range $100 \leq $~$h$~$ \leq 2500$~km shows a near $60$-fold decrease in evaluation time for $\rho$${}_S$ compared to $\rho$${}_J$.
The implementation of the Jacchia-77 model used herein is written in the coding language \codingC{}~\citep[taken from][]{Jacc2017}, and called from within \matlab, while the routine to calculate $\rho$${}_S$ is implemented and called directly in \matlab.
Thus, a further decrease of computational time could be expected if also the latter was implemented in \codingC.
The speed tests were performed using the same processor architecture.

\subsubsection{Variable Model}
\label{sec:variableAtmosphere}
Possible extensions to the smooth exponential atmosphere model are the inclusion of a temporal dependence, such as the solar cycle, annual or daily variations.
Here, the model is extended to incorporate the variability in the atmosphere density due to a variable $T_\infty$.
To conserve the mathematical formulation of the static model, the temperature dependence is introduced in the fitting parameters, $\hat{\rho}_p=\hat{\rho}_p(T_\infty)$ and $H_p=H_p(T_\infty)$. 

$T_\infty$ is a function of the solar proxy $F$ (see Equation~\ref{eq:Tinf}), so the fitting range is defined by $F$.
Generally, the long-term predictions for $F$ -- based on various numbers of previous solar cycles -- remain between $F \in [60, 230]$~sfu~\citep{Vallado2014, Perez2015, Radtke2016}.
This translates into $T_\infty \in [669, 1321]$~K, as $\overline{F}$ per definition remains in the same range as $F$.
The parameters for the variable smooth exponential atmosphere model derived below, and listed in ~\ref{sec:variableAtmosphereParameters}, are valid for any $T_\infty \in [T_0=650, T_1=1350]$~K. They should not be used for $T_\infty$ outside this range, as polynomial fits tend to oscillate strongly outside the fitting interval.

The dependence on $T_\infty$ is incorporated using a polynomial least squares fit.
Each partial atmosphere is fitted separately.
The static parameters, fitted to the $i=1, 2, \dots, M$ static atmospheres with different $T_\infty$, are converted
\begin{subequations}
	\label{eq:tologspace}
	\begin{align}
	a_p^i &= -1/H_p^i \\
	b_p^i &= \ln(\hat{\rho}_p^i)
	\end{align}
\end{subequations}
and each time-variable partial atmosphere is fitted to two independent polynomials of order $l$ and $m$ respectively
\begin{subequations}
	\label{eq:variablemodel}
	\begin{align}
	a_p(\tilde{T}_\infty) &= \sum_{k=0}^{l} a_{pk} \tilde{T}_\infty^k \\
	b_p(\tilde{T}_\infty) &= \sum_{k=0}^{m} b_{pk} \tilde{T}_\infty^k
	\end{align}
\end{subequations}
using a normalised and unit-less $\tilde{T}_\infty$, defined as
\begin{equation}
\tilde{T}_\infty = \frac{T_\infty-T_0}{T_1-T_0}
\end{equation}
In vector notation, Equation~\ref{eq:variablemodel} can be written as
\begin{subequations}
	\label{eq:abvectors}
	\begin{align}
	\pmb{a} &= \begin{bmatrix} a_1 \\ \vdots \\ a_{n_p} \end{bmatrix} = \begin{bmatrix} a_{10} & \dots & a_{1l} \\ \vdots & \ddots &  \\ a_{n_p 0} & \dots & a_{n_p l} \end{bmatrix} \begin{bmatrix} \tilde{T}_\infty^0 \\ \vdots \\ \tilde{T}_\infty^l \end{bmatrix} \\
	\pmb{b} &= \begin{bmatrix} b_1 \\ \vdots \\ b_{n_p} \end{bmatrix} = \begin{bmatrix} b_{10} & \dots & b_{1m} \\ \vdots & \ddots &  \\ b_{n_p 0} & \dots & b_{n_p m} \end{bmatrix} \begin{bmatrix} \tilde{T}_\infty^0 \\ \vdots \\ \tilde{T}_\infty^m \end{bmatrix}
	\end{align}
\end{subequations}
To prevent over-fitting, the order of the polynomials should remain well below the number of fitted static atmospheres.
Here, the model in Equation~\ref{eq:variablemodel} is fitted to $M=50$ statically fitted models, distributed again as Chebyshev nodes between $T_0$ and $T_1$
\begin{equation}
\label{eq:ChebyshevTemp}
	T_i = \frac{T_0+T_1}{2}+\frac{T_1-T_0}{2} \cos \left( \frac{2i-1}{2N} \pi \right) \qquad i=1, \dots, N
\end{equation}
The orders are chosen to be $l=m=8$ such that the error remains below $0.5\%$ for all $h > 155$~km and $T_\infty \in [650, 1350]$~K.
\begin{figure}
	\centering
	\includegraphics[width=1.\linewidth]{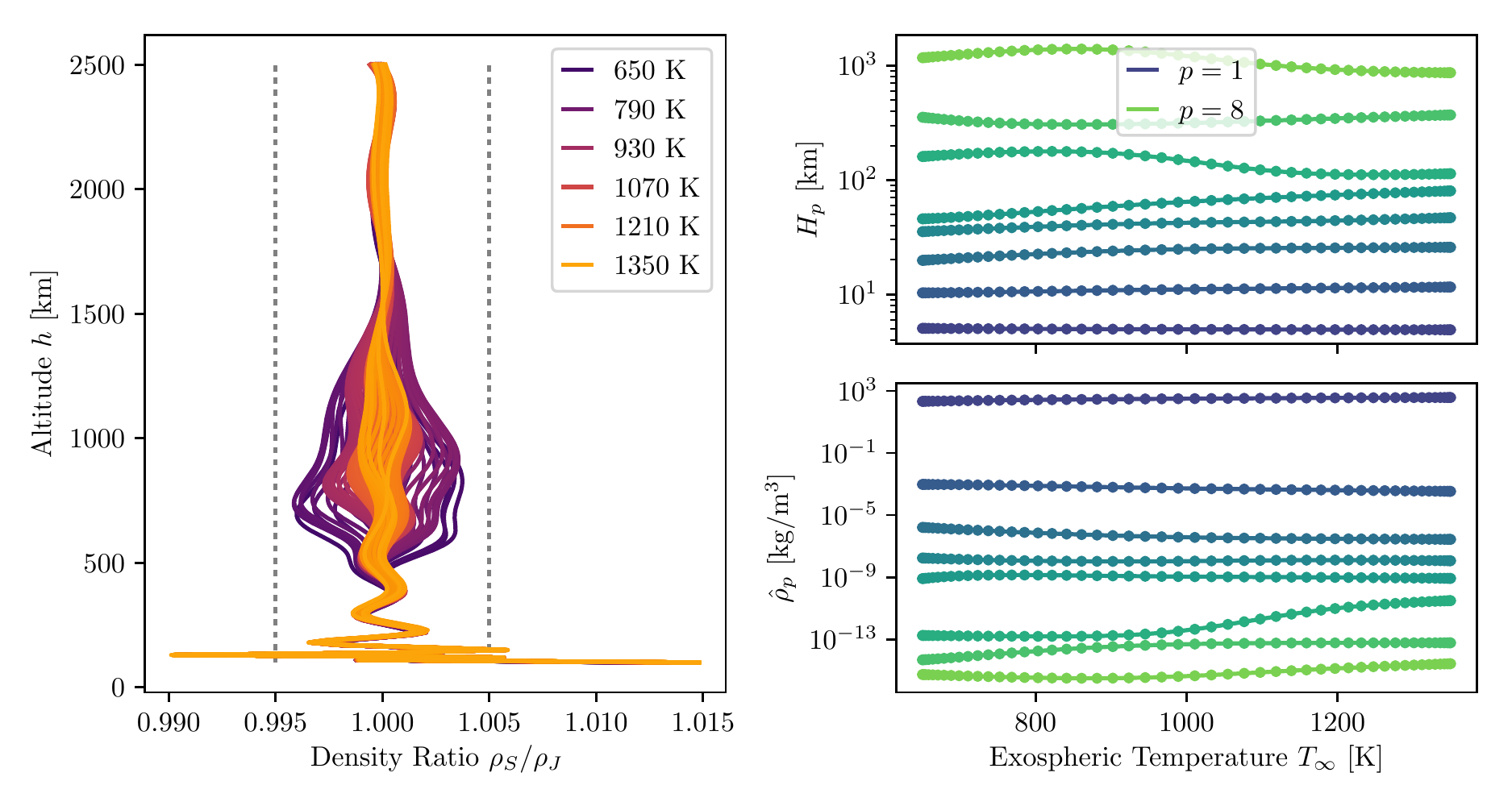}
	\caption{Quality of temperature dependent fit. Left: comparison for different $T_\infty$. Right: evolution of $H$${}_p$ (top) and $\hat{\rho}$${}_p$ (bottom) as a function of $T_\infty$. The dots show the parameters of the static fits, which were used to fit the variable model.}
	\label{fig:fitVariable}
\end{figure}
If more accuracy is needed, the polynomial order can be increased and/or spline polynomial interpolation applied.
Finally, the time-dependent atmosphere is recovered by inverting Equation~\ref{eq:tologspace}
\begin{subequations}
	\label{eq:varparams}
	\begin{align}
	H_p(T_\infty) &= -1/a_p (\tilde{T}_\infty) \\
	\hat{\rho}_p(T_\infty) &= \exp({b_p(\tilde{T}_\infty)})
	\end{align}
\end{subequations}
Figure~\ref{fig:fitVariable} compares the accuracy of the  $T_\infty$-variable smooth exponential atmosphere model against the original Jacchia-77 model.
It shows the ratio between $\rho_S(T_\infty)/\rho_J(T_\infty)$ for $T_\infty$ in the range from $650$~K to $1350$~K (left), and the corresponding parameters, $\hat{\rho}$${}_p$ and $H$${}_p$ as a function of $T_\infty$, including the underlying parameters of the static fits (right).
Towards the lower edge of the temperature range (i.e. $T_\infty \rightarrow 650$~K), the polynomial fits for components $p=5-7$ do not well represent the underlying data.
This leads to increased but still tolerable errors in the altitude range between $500$ and $1500$~km.

The advantage of this approach is, that the original structure of the model is maintained, so it can be used with the contraction model introduced in the next section.

\subsection{Superimposed King-Hele Approximation}
\label{sec:SIKH}
The extension of the KH contraction formulation into the SuperImposed King-Hele (SI-KH) formulation with a superimposed atmosphere is straightforward. 
Replacing $\rho$ from Equation~\ref{eq:averagedLagrangePlanetary} with the one defined in Equation~\ref{eq:smoothAtmosphere} leads to
\begin{subequations}
	\label{eq:SuperimposedContraction}
	\begin{align}
		\Delta a &= \sum_{p=1}^{n_p} \Delta a_p
				 = -a^2 \delta \sum_{p=1}^{n_p} \int_{0}^{2 \pi} \rho_p \frac{(1+e \cos{E})^{\frac{3}{2}}}{(1-e \cos{E})^{\frac{1}{2}}} \wrt{E} \\
		\Delta e &= \sum_{p=1}^{n_p} \Delta e_p
				 = -a \delta \sum_{p=1}^{n_p} \int_{0}^{2 \pi} \rho_p \left( \frac{1+e \cos{E}}{1-e \cos{E}} \right)^\frac{1}{2} \cos{E} (1-e^2) \wrt{E}
	\end{align}
\end{subequations}
i.e. each partial contraction reduces to the classical KH formulation with the partial exponential atmosphere $\rho$${}_p$. The important difference is that now $H$${}_p$ is constant over the whole altitude range.
The classical KH approximations -- extended up to 5\textsuperscript{th} order -- can be found in~\ref{sec:KH} (dropping the subscript $p$).
Finally, the rate of change is
\begin{equation}
	F_x = \dv{x}{t} = \sum_{p=1}^{n_p} \left( F_x \right)_p \approx \frac{1}{P} \sum_{p=1}^{n_p} \Delta x_p \qquad x \in [a, e]
\end{equation}

KH introduced the simple fixed boundary condition $e_b$$=0.2$ to select between the approximation method for low eccentric and high eccentric orbits, given in~\ref{sec:loweccentric} and~\ref{sec:higheccentric}, respectively. However, as $H$${}_p$  can be large, this condition is not always sufficient. Recall from Equation~\ref{eq:auxilaryvariable} that
\begin{equation}
	z=\frac{a e}{H}
\end{equation}	
For low $a$ and high $H$, $z$ can approach unity at $e=0.2$, making the series expansion in $1/z$ inaccurate.
Instead, it is proposed to define $e_b$ based on the truncation errors found in the formulations for the low and high eccentric orbits.
The series truncation errors for the low eccentric orbit approximation (Equation~\ref{eq:contLow}), using the order notation, $\mathcal{O}$, are of the order of
\begin{subequations}
	\label{eq:truncLow}
	\begin{align}
	\mathcal{O}_{a}^{low}(e^6) &= a^2 \rho \exp{(-z)} I_{0} e^6 \\
	\mathcal{O}_{e}^{low}(e^6) &= a \rho \exp{(-z)} I_{1} e^6
	\end{align}
\end{subequations}
If $z$ is large (see justification below), $I_{0/1}(z) \to \exp(z)/\sqrt{2 \pi z}$ and Equation~\ref{eq:truncLow} becomes
\begin{subequations}
	\label{eq:truncLowSimpl}
	\begin{align}
	\mathcal{O}_{a}^{low}(e^6) &= a^2 \rho \frac{e^6}{\sqrt{z}} \\
	\mathcal{O}_{e}^{low}(e^6) &= a \rho \frac{e^6}{\sqrt{z}}
	\end{align}
\end{subequations}
For the high eccentric orbit approximation (Equation~\ref{eq:contHigh}), the truncation errors are in the order of 
\begin{subequations}
	\label{eq:truncHigh}
	\begin{align}
	\mathcal{O}_{a}^{high}(\frac{1}{z^6}) &= a^2 \rho \frac{1}{\sqrt{z}} \frac{(1+e)^{\frac{3}{2}}}{(1-e)^{\frac{1}{2}}} \frac{1}{z^6(1-e^2)^6} \\
	\mathcal{O}_{e}^{high}(\frac{1}{z^6}) &= a \rho \frac{1}{\sqrt{z}} \left( \frac{1+e}{1-e} \right)^\frac{1}{2}\frac{1}{z^6 (1-e^2)^5}
	\end{align}
\end{subequations}
Assuming that the terms $\frac{(1+e)^{\frac{3}{2}}}{(1-e)^{\frac{1}{2}}}\frac{1}{(1-e^2)^6}$ and $\left( \frac{1+e}{1-e} \right)^\frac{1}{2}\frac{1}{(1-e^2)^5}$ are dominated by $1/z^6$ (see again below for a justification), Equation~\ref{eq:truncHigh} simplifies to
\begin{subequations}
	\label{eq:truncHighSimpl}
	\begin{align}
	\mathcal{O}_{a}^{high}(\frac{1}{z^6}) &= a^2 \rho \frac{1}{\sqrt{z}} \frac{1}{z^6} \\
	\mathcal{O}_{e}^{high}(\frac{1}{z^6}) &= a \rho \frac{1}{\sqrt{z}} \frac{1}{z^6}
	\end{align}
\end{subequations}
Equating the truncation errors from Equations~\ref{eq:truncLowSimpl} and~\ref{eq:truncHighSimpl}, using Equation~\ref{eq:auxilaryvariable} and solving for $e$ results in the following condition
\begin{equation}
	\label{eq:contBoundary}
	e_b = \sqrt{\frac{H}{a}}
\end{equation}
Note that this boundary is most exact if the series expansions in both the low and high eccentric regimes are of the same order.

The derivation of the boundary condition required the assumptions of $z$ to be large, such that $I_{0/1}(z) \to \exp(z)/\sqrt{2 \pi z}$ and such that $1/z^6$ dominates the other $e$ terms in Equation~\ref{eq:truncHigh}. To validate the assumptions, replace $a$ in Equation~\ref{eq:contBoundary} with $a=(h_p+R_E)/(1-e_b)$ and solve for $e_b$, neglecting the negative solution
\begin{equation}
	\label{eq:assumptionProof}
	\begin{aligned}
	e_b = \frac{1}{2}[-y+\sqrt{y^2+4y}] \qquad \textrm{where} \quad y=\frac{H}{h_p+R_E}
	\end{aligned}
\end{equation}
Given $H_{min/max}=4.9/1320$~km (see Table~\ref{tab:FitParameters}) and the valid range for $h_p \in h \in [100, 2500]$~km, the extrema in $e_b$ and $z_b=1/e_b$, are found to be
\begin{subequations}
	\begin{align*}
	e_{b, min/max} &= 0.023/0.361 \\
	z_{b, min/max} &= 2.77/43
	\end{align*}
\end{subequations}
For any $z > z_{b, min}$, $I_{0/1}$ remains close to $\exp(z)/\sqrt{2 \pi z}$, being off only $+6\%$ and $-16\%$, respectively, at $z_{b, min}$. At the same time, $1/z^6$ dominates the terms dependent on $e$ in Equation~\ref{eq:truncHigh} by two to three orders of magnitude $\forall e < e_{b, max}$.
Thus, the assumptions made to derive $e_b$ are valid.

An advantage of an analytical expression of the dynamics is that the Jacobian of the dynamics can be derived analytically too, which can be used for uncertainty propagation.
For a comprehensive discussion of the SI-KH method, the partial derivatives of the dynamics as derived by KH, with respect to $a$ and $e$, are given in \ref{sec:jacobian} (again dropping the subscript $p$).
As the SI-KH method is simply a summation of the individual contributions of the partial atmosphere, the derivatives can equally be summed up as
\begin{equation}
	\pdv{F_x}{y} = \sum_{p=1}^{n_p} \left( \pdv{F_x}{y} \right)_p \qquad (x, y) \in [a, e]
\end{equation}
\section{Validation}
The validation section is split into two parts: Section~\ref{sec:resultatmos} validates the smooth exponential atmosphere, $\rho$${}_S$, by comparing it to the Jacchia-77 model, $\rho$${}_J$, during SA propagation using the GL contraction method;
Section~\ref{sec:resultsingle} validates the proposed SI-KH approach by comparing the contraction approximation along a single orbit, i.e. $\Delta a$ and $\Delta e$, to numerical quadrature.
For completeness, propagations of a grid of initial conditions are performed using the GL and SI-KH methods and NA integration.
The latter does not resort to any averaging technique, integrating the full dynamics of Equation~\ref{eq:modLagrangePlanetary}, including~$E$.

\subsection{Validation of the Smooth Exponential Atmosphere Model}
\label{sec:resultatmos}
To validate $\rho$${}_S$ against $\rho$${}_J$ for $T_\infty$~$=750$, $1000$ and $1250$~K and at the same time distinguish it from the effects introduced by the SI-KH method on the resulting lifetime, $t_L$, the following orbits are propagated using the GL method only for the computation of the orbit contraction.
All physically feasible initial orbit configurations on a $46 \times 46$ grid from $250 \leq h_p \leq 2500$~km and $250 \leq h_a < 2500$~km are propagated, using $\delta=1$~m$^2$/kg.
The lower limit, $250$~km, is selected as an object with such a large $\delta$ on a circular orbit at this altitude survives for a fraction of a day only at which point SA propagation becomes inaccurate.
The upper limit, $2500$~km, is being imposed by definition of $\rho$${}_J$, but can be overcome by fitting to another model. 
The chosen $\delta$ is large, but does not limit the validity of this validation, as inaccuracies from the SA approach affect the propagation equally for both atmosphere models.

The SA propagation is performed using \matlab's \ode{113} -- a variable-step, variable-order Adams-Bashforth-Moulton integrator~\citep{Shampine1997} -- and a relative error tolerance, $\gamma_{rel}$$=10^{-6}$, which is shown to be sufficient for different orbital scenarios in Section~\ref{sec:resultsingle}.
\begin{table}
	\centering
	\caption{Comparison of $1081$ propagations being subject to $\rho$${}_J$ or $\rho$${}_S$, in total number of function evaluations, $N_f^{tot}$, total integration evaluation time, $t_{CPU}^{tot}$, and the minimum and maximum lifetime estimation error, $\eta_{t_L, min/max}$.}
	\begin{tabular}{ r l c r c } 
 \hline
 $T_\infty$ [K]	& $\rho$	& $N_f^{tot}$ [$-$]	& $t_{CPU}^{tot}$ [s]	& $\eta_{t_L, min/max}$ [$\%$]\\
 \hline
 $750$			& $\rho_J$	& $593255$			& $1086.9$				& \\
				& $\rho_S$	& $592124$			& $169.5$				& $-0.060/0.051$ \\
 $1000$			& $\rho_J$	& $568140$			& $986.3$				& \\
 				& $\rho_S$	& $568140$			& $153.6$				& $-0.077/0.056$ \\
 $1250$			& $\rho_J$	& $550021$			& $789.9$				& \\
 				& $\rho_S$	& $549063$			& $149.3$				& $-0.074/0.048$ \\
 \hline
\end{tabular}
	\label{tab:comparisonAtmosphere}
\end{table}
\begin{figure}
	\centering
	\subfloat[For low solar activity, i.e. $T_\infty=750$~K.]{
		\includegraphics[width=\linewidth]{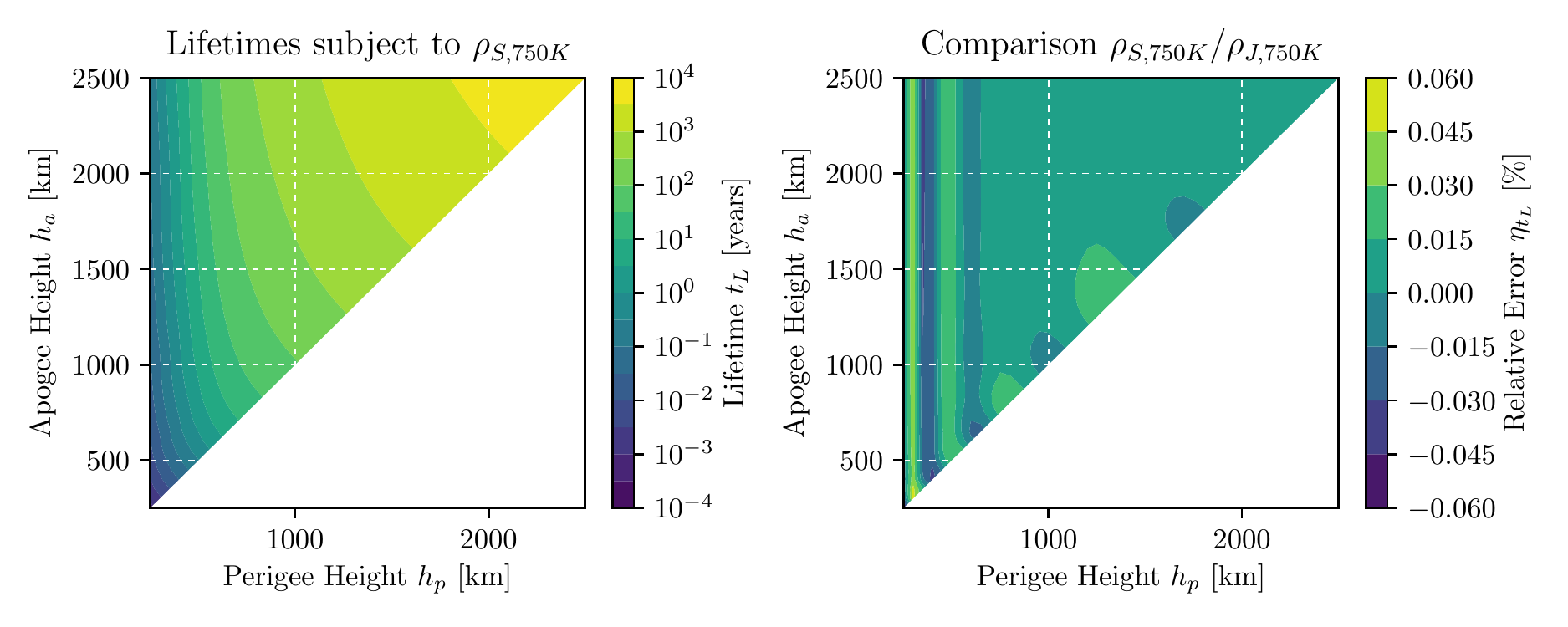}
		\label{fig:comparisonAtmosphereT750}}

	\subfloat[For mean solar activity, i.e. $T_\infty=1000$~K.]{
		\includegraphics[width=\linewidth]{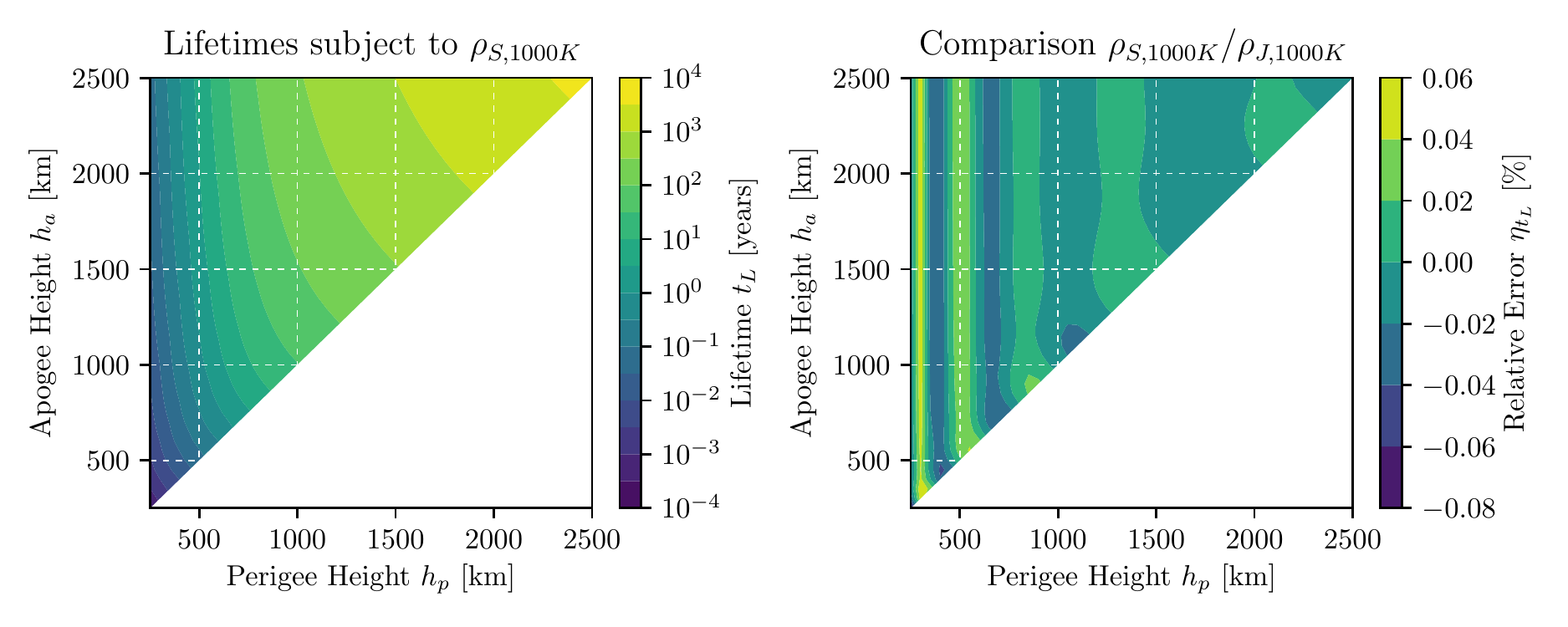}
		\label{fig:comparisonAtmosphereT1000}}
	
	\subfloat[For high solar activity, i.e. $T_\infty=1250$~K.]{
		\includegraphics[width=\linewidth]{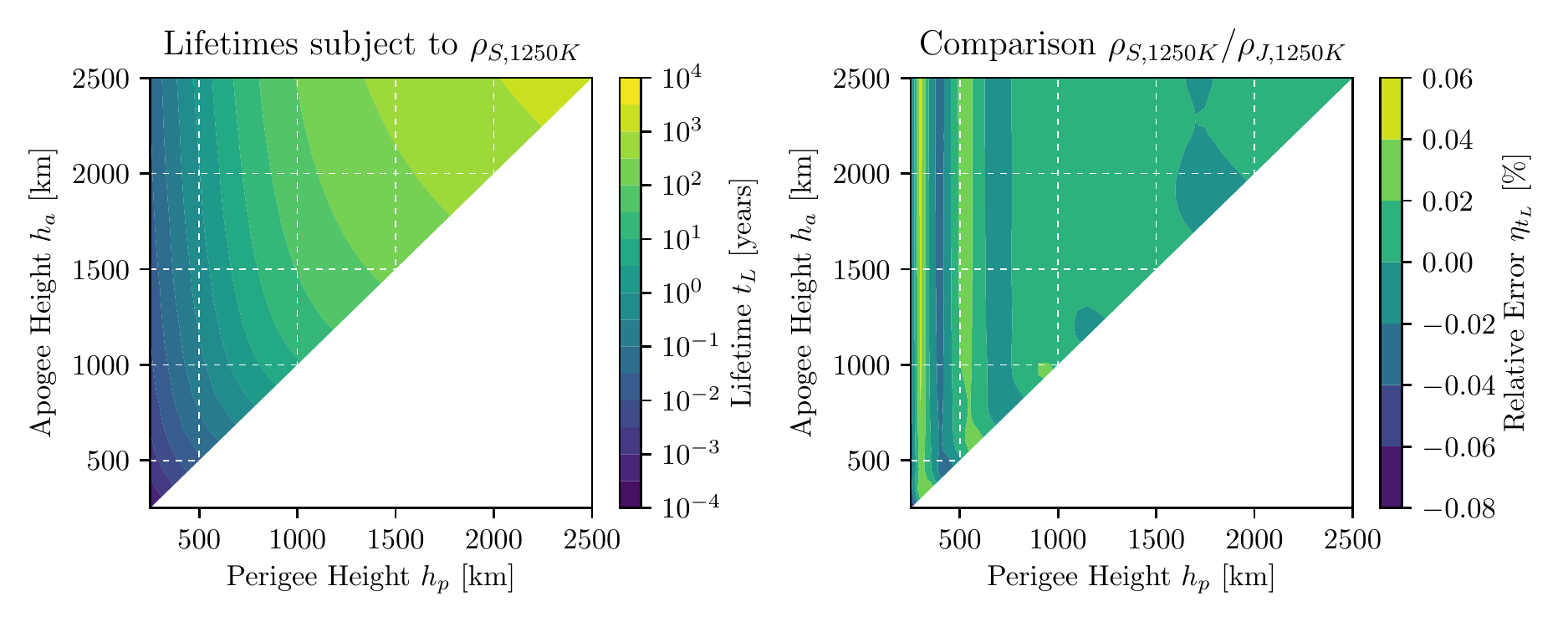}
		\label{fig:comparisonAtmosphereT1250}}
	\caption{Lifetimes and comparison of accuracy for lifetime estimation for objects being subject to $\rho$${}_J$ and $\rho$${}_S$.}
	\label{fig:comparisonAtmosphere}
\end{figure}
Figure~\ref{fig:comparisonAtmosphere} shows $t_L$ for the initial orbit grid, for propagation subject to $\rho$${}_S$, and the relative error,~$\eta_{t_L}$, defined as
\begin{equation}
	\eta_{t_L} = \frac{t_L(\rho_S)-t_L(\rho_J)}{t_L(\rho_J)}
\end{equation}
comparing the propagations for each grid point using $\rho$${}_S$ and $\rho$${}_J$, respectively.
Table~\ref{tab:comparisonAtmosphere} contains information about the maximum error and the workload.
Over the whole specified domain and for all $T_\infty$~$ \in [750, 1000, 1250]$~K, $\eta_{t_L}$ remains within $[-0.1\%, 0.1\%]$, which considering the uncertainties in atmospheric density modelling is more than accurate enough~\citep{Sagnieres2017}.
Towards low perigees ($h_p<500$~km), the fitted $\rho$${}_S$ starts to wobble around the underlying model (see Figure~\ref{fig:Fit1000Density}), which is also apparent for the propagated orbits.
A $6$-fold speed improvement can be observed, as no numerical integration is required when calculating the density with $\rho$${}_S$.

The reduction in function evaluations and computational time observable with an increasing $T_\infty$ is a consequence of the different density profiles.
Increasing $T_\infty$ leads to an increased $\rho$, which increases the drag force and thus decreases the lifetime.
However, the variable-step size integration method can compensate this by increasing the step size.
Two possible explanations are: as the integrator is initialised with the same properties for all three cases, the initially set (small) step size favours shorter lifetimes; and the shape of the density profiles with high $T_\infty$ are more smooth, decreasing the number of failed function evaluation attempts.

\subsection{Validation of the superimposed King-Hele method}
\label{sec:resultsingle}
\begin{figure}
	\centering
	\subfloat[Analytical KH approximation compared against GL quadrature. Differences of up to $25\%$ can occur for certain orbital configurations.]{
		\includegraphics[width=\linewidth]{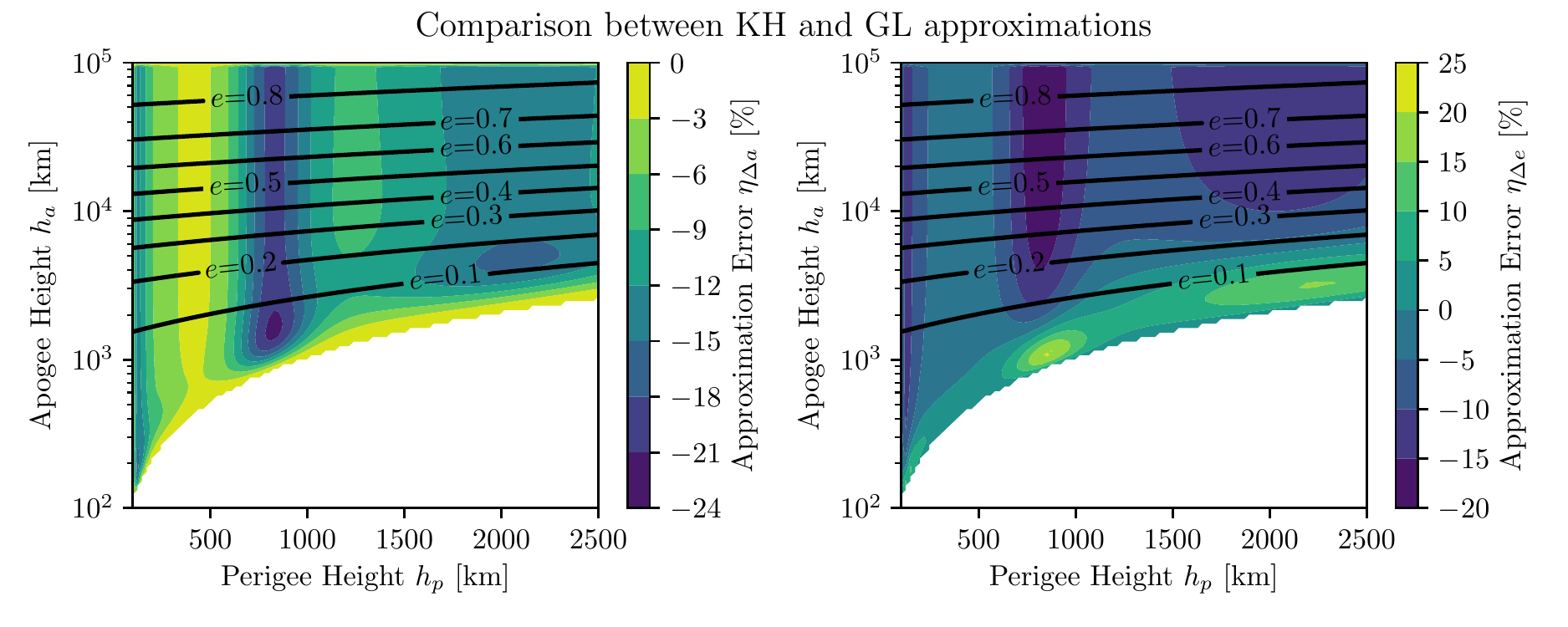}
		\label{fig:deltaComparisonKH}}

	\subfloat[Analytical SI-KH approximation compared against GL quadrature. The error remains below $0.1\%$ across the domain.]{
		\includegraphics[width=\linewidth]{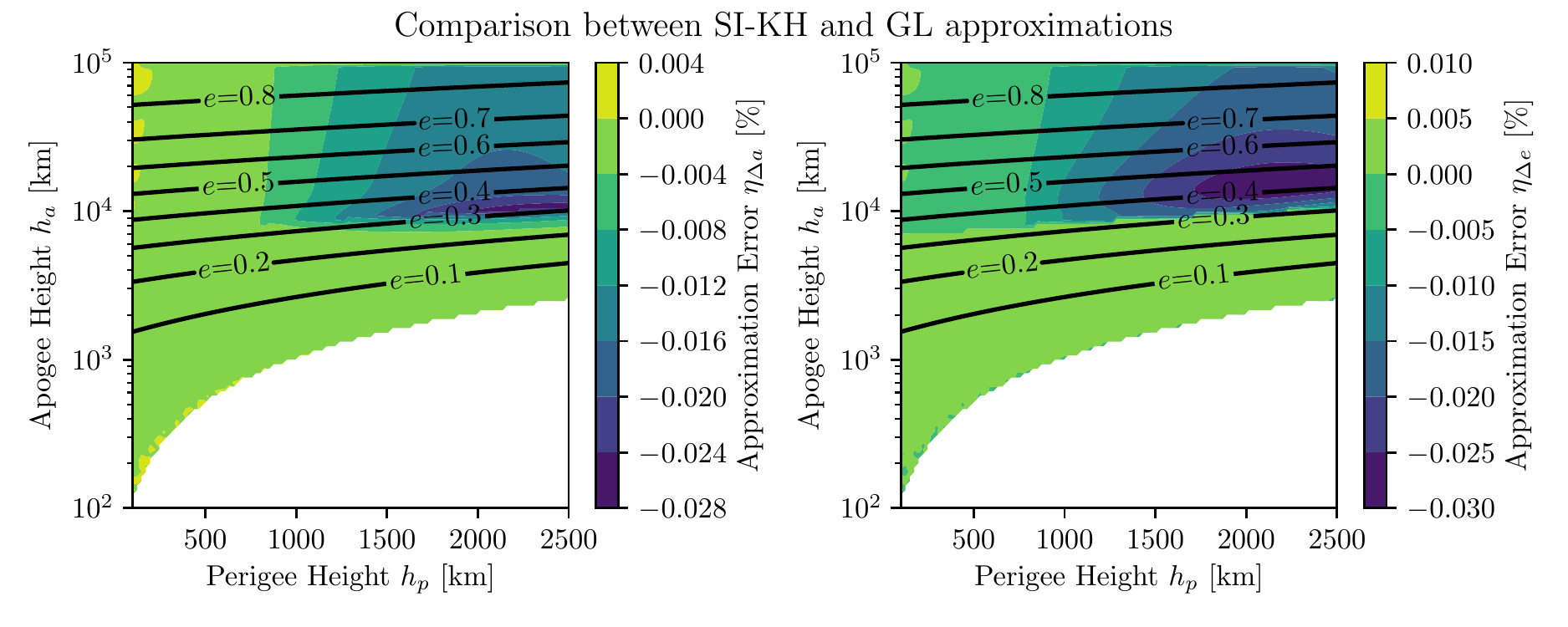}
		\label{fig:deltaComparisonSIKH}}
	\caption{Comparison for accuracy in $\Delta a$ (left) and $\Delta e$ (right) for different approximation methods. The underlying atmosphere model is $\rho$${}_S$ at $T_\infty=1000$~K. Note that the colour bar range of the lower figure is 3 orders of magnitudes smaller than the one of the upper figure. }
	\label{fig:deltaComparison}
\end{figure}

The SA propagation relies on an accurate approximation of $\Delta a$ and $\Delta e$.
Figure~\ref{fig:deltaComparison} shows -- for different orbital configurations -- the relative integral approximation error, $\eta_{\Delta x}$, defined as
\begin{equation}
	\eta_{\Delta x} = \frac{\Delta x(\mathcal{C}_2)-\Delta x(\mathcal{C}_1)}{\Delta x(\mathcal{C}_1)} \qquad x \in [a, e]
\end{equation}
where $\mathcal{C}$ is the selected contraction method: $\mathcal{C}_1$ is the numerical GL method computed using 65~nodes; and $\mathcal{C}_2$ describes the analytical formulation, KH or SI-KH, using series expansion up to $5$\textsuperscript{th} order.

Figure~\ref{fig:deltaComparisonKH} reveals why orbits are predicted to re-enter much later using the classical KH contraction method: the density is underestimated at altitudes above $h_p$.
The largest errors occur around $h_p=125$~km and $800$~km, where the rate of change in $H$ with respect to $h$ is large.
Around these two altitudes, the contraction rate in $a$ is underestimated by more than $10\%$ and $20\%$, respectively, if $e > 0.03$.
Using the SI-KH the relative error remains well below $0.1\%$ $\forall h_p \in [100, 2500]$~km and $\forall h_a \in [100, 100000]$~km (see Figure~\ref{fig:deltaComparisonSIKH}), a range that includes the vast majority of all Earth orbiting objects.
Discontinuities can be found whenever $e$ passes through $e_b=e_b(H_p)$.
The biggest step occurs for the largest $H$${}_p$.
Those discontinuities slightly increase the number of steps required during the integration.
However, given the averaged dynamics, $\gamma_{rel}$ can be chosen large enough during integration mitigating the effects of the discontinuities.

To see how the SI-KH compares against GL in SA propagation and against NA propagation in terms of accuracy and computational power, the results from different initial orbit conditions are compared, for two scenarios:
\begin{enumerate}[label=\alph*)]
	\item Short-term re-entry duration: $t_L=30$~days
	\item Mid-term re-entry duration: $t_L=360$~days
\end{enumerate}
The reasons why long-term re-entry cases are not discussed here are two-fold:
First, for long time spans, the NA integration requires small relative tolerances.
If they are not met, the result cannot be trusted; Secondly, the longer the time spans, i.e. the smaller $\delta$, the more accurate the assumptions made for the SA propagation.

\begin{figure}
	\centering
	\subfloat[Effective area-to-mass ratio required to re-enter in 30~days.]{
		\includegraphics[width=.47\linewidth]{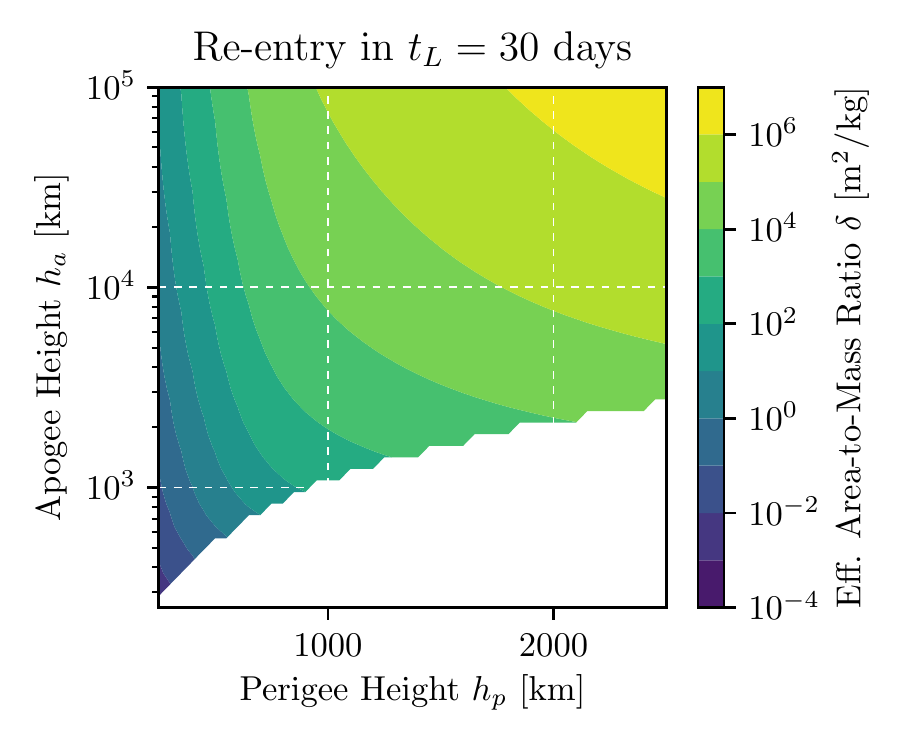}
		\label{fig:delta30Days}}
	\hfill
	\subfloat[Effective area-to-mass ratio required to re-enter in 360~days.]{
		\includegraphics[width=.47\linewidth]{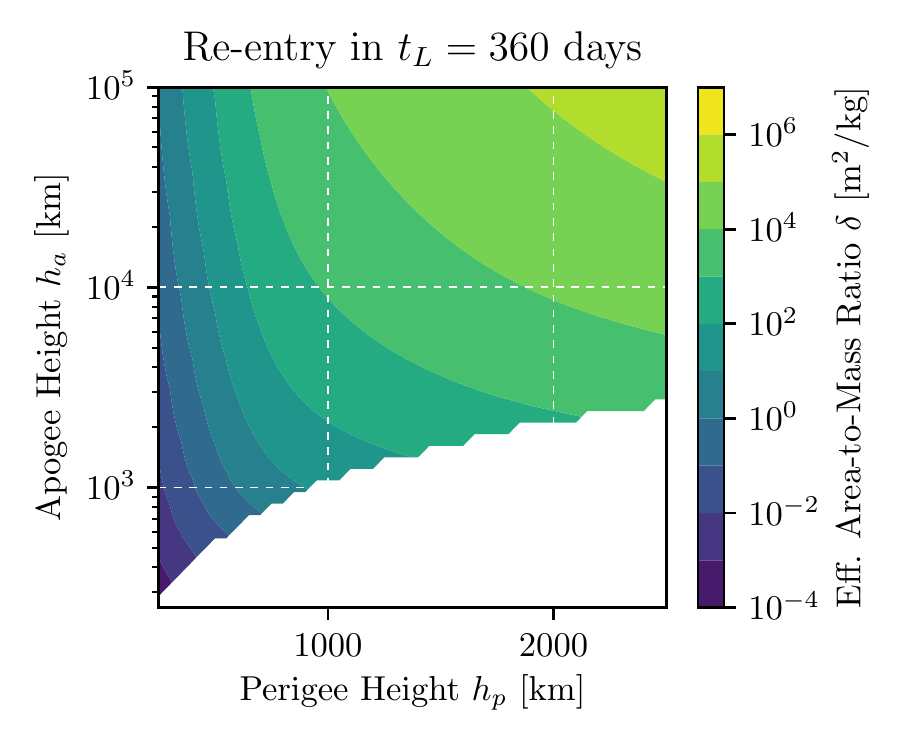}
		\label{fig:delta360Days}}
	\caption{The minimum effective area-to-mass ratio is $\delta_{min}=1.5 \times 10^{-4}$ in order to remain in orbit for 360~days from a circular orbit at $h_p=h_a=250$~km. The maximum, in order to re-enter in $30$~days from $h_p/h_a=250/100000$~km, is $\delta_{max}=3.0 \times 10^6$.}
	\label{fig:deltaDays}
\end{figure}

The initial conditions are spaced in $h_p \in [250, 2500]$~km and $h_a \in [250, $ $100000]$~km and consist of all the $1558$ feasible solutions on a $46 \times 46$ grid, where the grid spacing in $h_a$ is chosen to be logarithmic, as opposed to the equidistant grid in $h_p$.
Two preliminary runs were performed using the SI-KH method to calculate the lifetimes. This way, the $\delta$ required to re-enter within the given time-span can be estimated.
Figure~\ref{fig:deltaDays} shows the grids of the resulting $\delta$ for both scenarios.
Note that $\delta$ varies by almost 13 orders of magnitude.

The accuracy is described again as the relative lifetime, $\eta_{t_L}$, this time defined as
\begin{equation}
	\eta_{t_L}^{ij} (\mathcal{M}_1, \mathcal{M}_2, h_{p, i}, h_{a, j}) = \frac{t_L(\mathcal{M}_1, h_{p, i}, h_{a, j})-t_L(\mathcal{M}_2, h_{p, i}, h_{a, j})}{t_L(\mathcal{M}_2, h_{p, i}, h_{a, j})}
\end{equation}
where $\mathcal{M}$ is the selected contraction and integration method, combined with a given relative integrator tolerance, $\gamma_{rel}$, during integration.
To give a feeling for the accuracy across all the different initial conditions, the $50\%$- and $100\%$-quantiles, i.e. the median and maximum denoted as $\eta_{t_L, 50\%}$ and $\eta_{t_L, 100\%}$, respectively, over all the $|\eta_{t_L}^{ij}|$ are given.
The computation effort is compared via the total number of function calls, $N_f^{tot}$, and time required for the integration itself, $t_{CPU}^{tot}$
\begin{subequations}
	\begin{align}
		\epsilon_{N_f}(\mathcal{M}_1, \mathcal{M}_2) &= \frac{N_f^{tot}(\mathcal{M}_1)}{N_f^{tot}(\mathcal{M}_2)} \\
		\epsilon_{t_{CPU}}(\mathcal{M}_1, \mathcal{M}_2) &= \frac{t_{CPU}^{tot}(\mathcal{M}_1)}{t_{CPU}^{tot}(\mathcal{M}_2)}
	\end{align}
\end{subequations}

\begin{table}
	\centering
	\caption{Performance of the different propagation and contraction methods, for a) $t_L=30$~days and b) $t_L=360$~days and various relative integration tolerances, $\gamma_{rel}$. All figures are unit less.}
	\begin{tabular}{ l l l | c c c c } 
 \hline
 	& $\mathcal{M}_1$ & $\mathcal{M}_2$	& $\eta_{t_L, 50\%}$ 	& $\eta_{t_L, 100\%}$ & $\epsilon_{N_f}$ & $\epsilon_{t_{CPU}}$ \\
 \hline
 a)	& SI-KH/$10^{-6}$	& SI-KH/$10^{-12}$	& $3.2\text{e}-6$	& $8.4\text{e}-5$	& $3.0\text{e}-1$	& $2.9\text{e}-1$	\\
 	& GL/$10^{-6}$	& GL/$10^{-12}$	& $3.3\text{e}-6$	& $7.0\text{e}-5$	& $3.7\text{e}-1$	& $3.7\text{e}-1$	\\
 	& NA/$10^{-6}$	& NA/$10^{-12}$	& $1.3\text{e}-3$	& $2.5\text{e}-2$	& $3.4\text{e}-1$	& $3.4\text{e}-1$	\\
 	& NA/$10^{-9}$	& NA/$10^{-12}$	& $1.6\text{e}-6$	& $3.1\text{e}-5$	& $6.0\text{e}-1$	& $6.2\text{e}-1$	\\
 	& SI-KH/$10^{-6}$	& NA/$10^{-12}$	& $8.7\text{e}-4$	& $1.8\text{e}-3$	& $1.1\text{e}-2$	& $2.2\text{e}-2$	\\
 	& GL/$10^{-6}$	& NA/$10^{-12}$	& $8.7\text{e}-4$	& $1.7\text{e}-3$	& $1.0\text{e}-2$	& $3.6\text{e}-2$	\\
 \hline
 b)	& SI-KH/$10^{-6}$	& SI-KH/$10^{-12}$	& $3.2\text{e}-6$	& $6.9\text{e}-5$	& $3.1\text{e}-1$	& $3.2\text{e}-1$	\\
 	& GL/$10^{-6}$	& GL/$10^{-12}$	& $3.7\text{e}-6$	& $6.9\text{e}-5$	& $3.9\text{e}-1$	& $4.0\text{e}-1$	\\
 	& NA/$10^{-6}$	& NA/$10^{-12}$	& $1.6\text{e}-2$	& $2.6\text{e}-1$	& $3.4\text{e}-1$	& $3.7\text{e}-1$	\\
 	& NA/$10^{-9}$	& NA/$10^{-12}$	& $1.9\text{e}-5$	& $4.1\text{e}-4$	& $6.1\text{e}-1$	& $6.4\text{e}-1$	\\
 	& SI-KH/$10^{-6}$	& NA/$10^{-12}$	& $7.0\text{e}-5$	& $3.2\text{e}-4$	& $5.8\text{e}-4$	& $1.1\text{e}-3$	\\
 	& GL/$10^{-6}$	& NA/$10^{-12}$	& $7.2\text{e}-5$	& $4.9\text{e}-4$	& $5.8\text{e}-4$	& $2.1\text{e}-3$	\\
 \hline
\end{tabular}
	\label{tab:performance}
\end{table}

Table~\ref{tab:performance} contains these figures comparing the different integration methods against each other.
For both SI-KH and GL, the absolute maximum error over the whole grid and over both scenarios remains below $0.01$\%, when decreasing $\gamma_{rel}$ from $10^{-6}$ to $10^{-12}$.
Given this force model, it is therefore sufficient to use $\gamma_{rel}=10^{-6}$.
For NA integration, this is not the case.
While the maximum error remains modest ($0.18\%$) in the short-term case, it becomes large when the re-entry span is increased to one year ($26\%$), when decreasing $\gamma_{rel}$.
Decreasing $\gamma_{rel}=10^{-9}$ and comparing to integration with $\gamma_{rel}=10^{-12}$, reduces the maximum error for the NA propagation in the mid-term case to $0.032\%$.

\begin{figure}
	\centering
	\subfloat[Small errors occur for large effective area-to-mass ratios ($\delta > 10^4$~m$^2$/kg).]{
		\includegraphics[width=.47\linewidth]{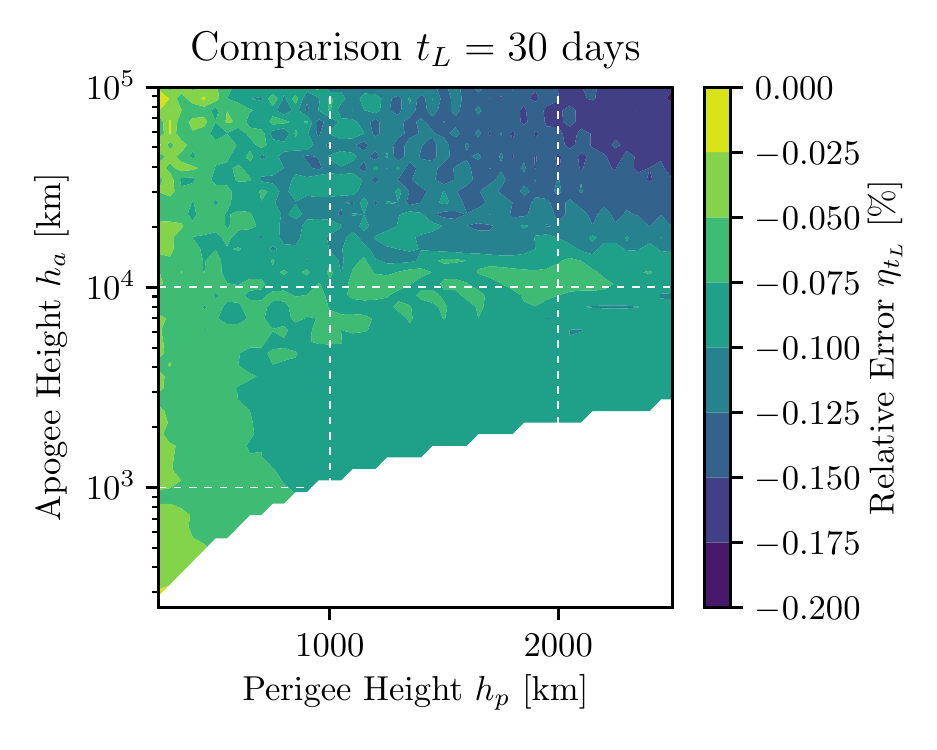}
		\label{fig:comparisonSIKHvsNum30days}}
	\hfill
	\subfloat[Two areas of very small errors can be distinguished, stemming from the series truncation.]{
		\includegraphics[width=.47\linewidth]{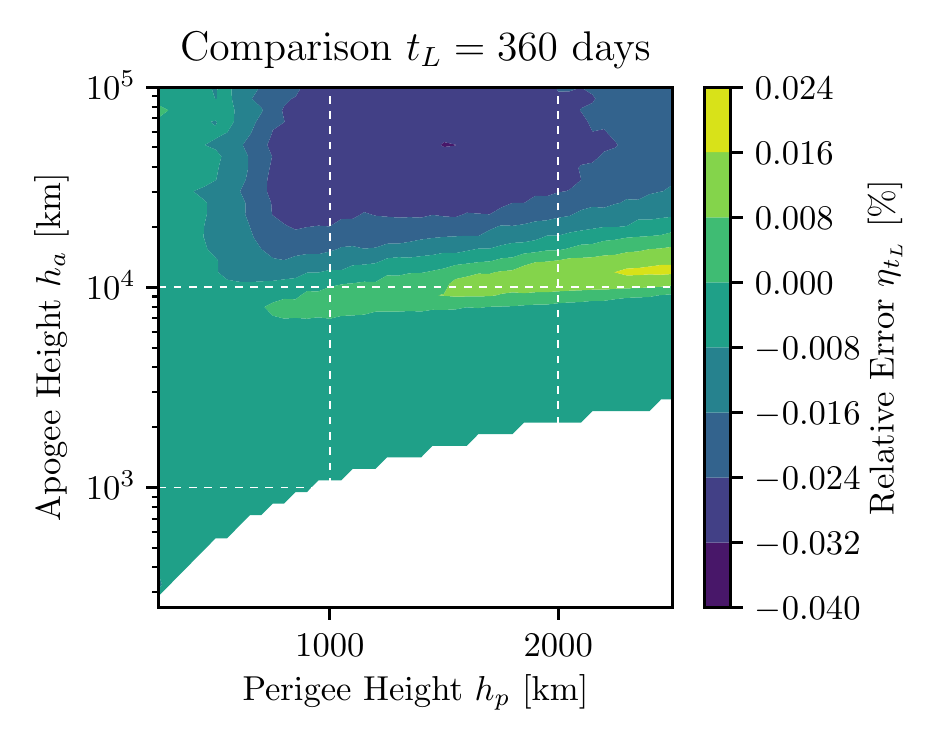}
		\label{fig:comparisonSIKHvsNum360days}}
	\caption{Relative error $\eta_{t_L}$ when comparing SA propagation using SI-KH with $\gamma_{rel}$$=10^{-6}$ against NA integration with $\gamma_{rel}$$=10^{-12}$.}
	\label{fig:comparisonSIKHvsNum}
\end{figure}

For the comparison of the SA techniques against NA propagation, the tolerance of the latter is set to $\gamma_{rel}=10^{-12}$.
Again, SI-KH and GL fare very similar.
For the short-term case, the boundaries of the SA propagation can be recognised for very high $\delta$, leading to still small maximum errors of $0.18\%$ and $0.17\%$, respectively. Figure~\ref{fig:comparisonSIKHvsNum30days} shows the resulting lifetime comparison for SI-KH and $t_L=30$~days.
As $\delta$ increases to values above $10^{4}$~m$^2$/kg, the assumption of constant $a$ and $e$ over one orbit starts to break down and small errors are introduced.
This might be an issue for small debris such as multi-layer insulation fragments and paint flakes.
For the mid-term scenario, the maximum error reduces by one order of magnitude for both SA methods tested.
For high $h_a > 10000$~km, the series expansion applied in the SI-KH method introduces small errors (see Figure~\ref{fig:comparisonSIKHvsNum360days}).
\section{Conclusion}
The classical KH orbit contraction method allows to analytically calculate the effects of drag on the orbit evolution averaged over an orbital period.
However, it inaccurately estimates the orbital decay for eccentric orbits subject to a non-exponentially decaying atmosphere model.
To improve the accuracy, a smooth exponential atmosphere model was proposed to be used in tandem with the new SI-KH orbit contraction method.

The classical KH method was extended to the SI-KH contraction method, making use of a superimposed atmosphere model to satisfy the assumption of a strictly decaying density for each component of the model.
This greatly reduces the errors in the estimated decay rates of objects in eccentric orbits and subject to atmospheric density profiles with variable scale height.
The analytical method was validated against an averaging technique based on numerical quadrature.
Further, the semi-analytical propagation of orbits using the SI-KH method was validated against full numerical integration of the dynamics.
The approach is applicable to any averaging techniques considering drag and based on the fixed scale height assumption above perigee.
Finally, the Jacobian of the dynamics governed by the SI-KH method is given to be used for future applications such as uncertainty propagation.
\section{Acknowledgements}
The research leading to these results has received funding from the European Research Council (ERC) under the European Union’s Horizon 2020 research and innovation programme as part of project COMPASS (Grant agreement No 679086).
The authors acknowledge the use of the Milkyway High Performance Computing Facility, and associated support services at the Politecnico di Milano, in the completion of this work.
The datasets generated for this study can be found in the repository at the link \href{http://www.compass.polimi.it/publications}{www.compass.polimi.it/ publications}.

\section{References}
\bibliographystyle{agsm}
\bibliography{database/bibliography}{}

\appendix
\section{Nomenclature}
\label{sec:nomenclature}
\begin{longtable}{l p{12cm}}
	
	$E$ & Eccentric anomaly [rad or deg] \\
	$F$ & $10.7$~cm solar flux [sfu] \\
	$\overline{F}$ & Smoothed $10.7$~cm solar flux [sfu] \\
	$H$ & Atmosphere density scale height [m or km] \\
	$I_n$ & Modified Bessel function of the first kind of order $n$ [$-$] \\
	$P$ & Orbit period [s] \\
	$R$ & Mean Earth radius [m or km] \\
	$T$ & Temperature [K] \\
	$T_\infty$ & Exospheric temperature [K] \\
	
	$\Delta a$ & Contraction over a full orbit period in a [m or km] \\
	$\Delta e$ & Contraction over a full orbit period in e [$-$] \\
	\\
	
	$a$ & Semi-major axis [m or km] \\
	$e$ & Eccentricity [$-$] \\
	$e_b$ & Boundary in e for selection of integral approximation method [$-$] \\
	
	$h$ & Height above Earth surface [m or km] \\
	$h_a$ & Apogee altitude [m or km] \\
	$h_p$ & Perigee altitude [m or km] \\
	
	$n_p$ & Number of partial atmospheres [$-$] \\
	$r$ & Radial distance from Earth's center [m or km] \\
	$t$ & Time [seconds, days or years] \\
	$t_L$ & Lifetime [seconds, days or years] \\
	$v$ & Intertial velocity [m/s or km/s] \\
	$z$ & Auxilary variable for integration of the decay rate of highly eccentric orbits [$-$] \\
	
	$\delta$ & Inverse ballistic coefficient [m$^2$/kg] \\
	$\eta_\rho$ & Relative atmospheric density error [$-$ or $\%$] \\
	$\eta_{\Delta x}$ & Relative integral approximation error [$-$ or $\%$] \\
	$\eta_{t_L}$ & Relative lifetime error [$-$ or $\%$] \\
	
	$\gamma_{rel}$ & Relative integration tolerance [$-$ or $\%$] \\
	$\mu$ & Earth gravitational parameter [m$^3$/s$^2$ or km$^3$/s$^2$] \\
	$\rho$ & Atmosphere density [kg/m$^3$ or kg/km$^3$] \\
	$\hat{\rho}$ & Atmosphere base density [kg/m$^3$ or kg/km$^3$] \\
	\\
	
	$\mathcal{C}$ & Contraction method \\
	$\mathcal{M}$ & Contraction and integration method with given $\gamma_{rel}$ \\
	$\mathcal{O}$ & Order of series truncation error \\
	\\
	
	${}_J$ & Index corresponding to the Jacchia-77 atmosphere model \\
	${}_S$ & Index corresponding to the smooth atmosphere model \\
	${}_p$ & Index corresponding to the partial smooth atmosphere model \\
	${}_{NS}$ & Index corresponding to the non-smooth atmosphere model
	
\end{longtable}

\section{King-Hele Formulation}
\label{sec:KH}
All the formulas presented here are explained and derived in the work of~\citet{King1964}. 
The analytical formulas describe, for different eccentricities, the change in the semi-major axis, $a$, and the eccentricity, $e$, over one orbit as an approximation of Equation~\ref{eq:averagedLagrangePlanetary}.
Please note that one of the four cases was dropped, as it was introduced only due to the Bessel functions becoming inaccurate for small arguments.
Today, the relevant mathematical software packages are accurate and fast enough to overcome this limitation.

Adaptations to the original formulation were made to 
\begin{itemize}
	\item find the change directly in $a$ and $e$, rather than $a$ and $x=ae$, to calculate the change in the variables of interest;
	\item find a more appropriate boundary condition, $e_b$, for the selection of the phase (see Section~\ref{sec:SIKH});
	\item increase the accuracy within each phase by taking into account more terms after the series expansion.
\end{itemize}

The two functions, $k_a$ and $k_e$, are introduced here for later use when describing the rate of change in all the eccentricity regimes described below, as
\begin{subequations}
	\label{eq:terms}
	\begin{align}
	k_a &= \delta \sqrt{\mu a} \rho(h_p) \\
	k_e &= k_a/a
	\end{align}
\end{subequations}
with the effective area-to-mass ratio, $\delta$, the gravitational parameter, $\mu$, the atmospheric density, $\rho$, evaluated at the perigee altitude, $h_p$.

\subsection{Circular Orbit}
For circular orbits, no integration needs to be approximated, as the integral can be solved analytically as
\begin{subequations}
	\label{eq:contCirc}
	\begin{align}
		\Delta a &= -2 \pi \delta a^2 \rho(h_p) \\
		\Delta e &= 0
	\end{align}
\end{subequations}
where $h_p$ reduces to the circular altitude. Dividing by the orbital period, $P$, according to Equation~\ref{eq:rateofchange} and using the functions defined in Equation~\ref{eq:terms}, the rate of change for circular orbits is
\begin{subequations}
	\label{eq:contCircDer}
	\begin{align}
		F_a = \dv{a}{t} &= -k_a \\
		F_e = \dv{e}{t} &= 0
	\end{align}
\end{subequations}

\subsection{Low Eccentric Orbit}
\label{sec:loweccentric}
For small $e < e_b(a, H)$, a series expansion in $e$ is performed and then integrated using the modified Bessel function of the first kind, $I_n$($z$), as
\begin{subequations}
	\label{eq:contLow}
	\begin{align}
			\pmb{e}^T	 &= 
				\begin{pmatrix}
					1 & e & e^2 & e^3 & e^4 & e^5
				\end{pmatrix} \notag \\
			\pmb{I}^T &= 
				\begin{pmatrix}
					I_0 & I_1 & I_2 & I_3 & I_4 & I_5 & I_6
				\end{pmatrix} \notag \\
			\Delta a &= -2 \pi \delta \rho(h_p) \exp{(-z)} a^2 [\pmb{e}^T \pmb{K}_a^l \pmb{I} + \mathcal{O}(e^6)] \\
			\Delta e &= -2 \pi \delta \rho(h_p) \exp{(-z)} a [\pmb{e}^T \pmb{K}_e^l \pmb{I} + \mathcal{O}(e^6)]
	\end{align}
\end{subequations}
with the auxiliary variable $z=a e / H$, the scale height, $H$, a single evaluation of the density at the perigee height, $h_p$, and the order of the series truncation error, $\mathcal{O}$, of $e^6$.
The constant matrices are given as
\begin{subequations}
	\begin{align*}
	\pmb{K}_a^l	&=
	\begin{bmatrix*}[r]
		1				& 0				& 0 			& 0				& 0				& 0				& 0 \\
		0				& 2				& 0 			& 0				& 0				& 0				& 0 \\
		\frac{3}{4}		& 0				& \frac{3}{4}	& 0				& 0				& 0				& 0 \\
		0				& \frac{3}{4}	& 0				& \frac{1}{4}	& 0				& 0				& 0 \\
		\frac{21}{64}	& 0				& \frac{28}{64}	& 0				& \frac{7}{64}	& 0				& 0 \\
		0				& \frac{30}{64}	& 0				& \frac{15}{64}	& 0				& \frac{3}{64}	& 0
	\end{bmatrix*} \\
	\pmb{K}_e^l &=
	\begin{bmatrix*}[r]
		0				& 1					& 0 				& 0					& 0				& 0				& 0 \\
		\frac{1}{2}		& 0					& \frac{1}{2}		& 0					& 0				& 0				& 0 \\
		0				& -\frac{5}{8}		& 0					& \frac{1}{8}		& 0				& 0				& 0 \\
		-\frac{5}{16}	& 0					& -\frac{4}{16}		& 0					& \frac{1}{16}	& 0				& 0	\\
		0				& -\frac{18}{128}	& 0					& -\frac{1}{128}	& 0				& \frac{3}{128}	& 0 \\
		-\frac{18}{256}	& 0					& -\frac{19}{256}	& 0					& \frac{2}{256}	& 0				& \frac{3}{256}	\\
	\end{bmatrix*}
	\end{align*}
\end{subequations}
Dividing by $P$ according to Equation~\ref{eq:rateofchange} and using the functions defined in Equation~\ref{eq:terms}, the rate of change for low eccentric orbits is
\addtocounter{equation}{-1}
\begin{subequations}
	\label{eq:contLowDer}
	\begin{align}
		F_a = \dv{a}{t}  &= -k_a \exp{(-z)} [\pmb{e}^T \pmb{K}_a^l \pmb{I} + \mathcal{O}(e^6)] \\
		F_e = \dv{e}{t}  &= -k_e \exp{(-z)} [\pmb{e}^T \pmb{K}_e^l \pmb{I} + \mathcal{O}(e^6)]
	\end{align}
\end{subequations}

\subsection{High Eccentric Orbit}
\label{sec:higheccentric}
Instead of performing the series expansion in $e$, which is infeasible for large values of $e > e_b(a, H)$, the expansion is performed for the substitute variable, $\lambda^2/z=1-\cos E$. KH truncated the series already after two powers. Here, instead, as $H$ can be large and as the formulation should be readily available for any $h_p < 2500$~km, it is extended up to 5\textsuperscript{th} power.
The contractions over one orbit period are
\begin{subequations}
	\label{eq:contHigh}
	\begin{align}
		\pmb{r}^T	 &= 
		\begin{pmatrix}
		1  & \frac{1}{z (1-e^2)}  & \frac{1}{z^2 (1-e^2)^2}  & \frac{1}{z^3 (1-e^2)^3} & \frac{1}{z^4 (1-e^2)^4}	& \frac{1}{z^5 (1-e^2)^5}
		\end{pmatrix} \notag \\
		\pmb{e}^T	 &= 
			\begin{pmatrix}
				1 & e & e^2 & e^3 & e^4 & e^5 & e^6 & e^7 & e^8 & e^9 & e^{10} \\
			\end{pmatrix} \notag \\
		\Delta a &= -2 \delta \sqrt{\frac{2 \pi}{z}} \rho(h_p) a^2 \frac{(1+e)^\frac{3}{2}}{(1-e)^\frac{1}{2}} [\pmb{r}^T \pmb{K}_a^h \pmb{e} + \mathcal{O}\left(\frac{1}{z^6}\right)] \\
		\Delta e &= -2 \delta \sqrt{\frac{2 \pi}{z}} \rho(h_p) a \left( \frac{1+e}{1-e} \right) ^\frac{1}{2} (1-e^2) [\pmb{r}^T \pmb{K}_e^h \pmb{e} + \mathcal{O}\left(\frac{1}{z^6}\right)]
	\end{align}
\end{subequations}
with the constant matrices
\begin{equation*}
	\pmb{K}_a^h =
	\begin{bmatrix*}[r]	
		\frac{1}{2}	& \frac{1}{16}	& \frac{9}{256}		& \frac{75}{2048}	& \frac{3675}{65536}	& \frac{59535}{524288} \\[.4em]
		0			& -\frac{1}{2}	& -\frac{3}{16}		& -\frac{45}{256}	& -\frac{525}{2048}		& -\frac{33075}{65536} \\[.4em]
		0			& \frac{3}{16}	& \frac{75}{128}	& \frac{675}{2048}	& \frac{5985}{16384}	& \frac{288225}{524288} \\[.4em]
		0			& 0				& \frac{3}{16}		& -\frac{75}{128}	& -\frac{105}{2048}		& \frac{10395}{16384} \\[.4em]
		0			& 0				& -\frac{15}{256}	& -\frac{3735}{2048}& \frac{21945}{32768}	& -\frac{344925}{262144} \\[.4em]
		0			& 0				& 0					& -\frac{45}{256}	& \frac{13545}{2048}	& -\frac{129465}{32768} \\[.4em]
		0			& 0				& 0					& \frac{105}{2048}	& \frac{110985}{16384}	& -\frac{7687575}{262144} \\[.4em]
		0			& 0				& 0					& 0					& \frac{525}{2048}		& -\frac{836325}{16384} \\[.4em]
		0			& 0				& 0					& 0					& -\frac{4725}{65536}	& -\frac{16288965}{524288} \\[.4em]
		0			& 0				& 0					& 0					& 0						& -\frac{33075}{65536} \\[.4em]
		0			& 0				& 0					& 0					& 0						& \frac{72765}{524288}				
	\end{bmatrix*}
\end{equation*}
\begin{equation*}
	\pmb{K}_e^h =
	\begin{bmatrix*}[r]
		\frac{1}{2}	& -\frac{3}{16}	& -\frac{15}{256}	& -\frac{105}{2048}	& -\frac{4725}{65536}	& -\frac{72765}{524288} \\[.4em]
		0			& -\frac{1}{4}	& \frac{9}{32}		& \frac{75}{512}	& \frac{735}{4096}		& \frac{42525}{131072} \\[.4em]
		0			& \frac{3}{16}	& \frac{39}{128}	& -\frac{405}{2048}	& \frac{525}{16384}		& \frac{152145}{524288} \\[.4em]
		0			& 0				& \frac{3}{32}		& -\frac{375}{256}	& \frac{735}{4096}		& -\frac{31185}{32768} \\[.4em]
		0			& 0				& -\frac{15}{256}	& -\frac{1515}{2048}& \frac{123585}{32768}	& -\frac{530145}{262144} \\[.4em]
		0			& 0				& 0					& -\frac{45}{512}	& \frac{31605}{4096}	& -\frac{1165185}{65536} \\[.4em]
		0			& 0				& 0					& \frac{105}{2048}	& \frac{40845}{16384}	& -\frac{10235295}{262144} \\[.4em]
		0			& 0				& 0					& 0					& \frac{525}{4096}		& -\frac{1505385}{32768} \\[.4em]
		0			& 0				& 0					& 0					& -\frac{4725}{65536}	& -\frac{5716305}{524288} \\[.4em]
		0			& 0				& 0					& 0					& 0						& -\frac{33075}{131072} \\[.4em]
		0			& 0				& 0					& 0					& 0						& \frac{72765}{524288}				
	\end{bmatrix*}
\end{equation*}
Plugging Equation~\ref{eq:contHigh} into Equation~\ref{eq:rateofchange}, using the functions defined in Equation~\ref{eq:terms}, and introducing the new functions
\begin{subequations}
	\label{eq:termsHigh}
	\begin{align}
	c_a &= \sqrt{\frac{2}{\pi z}} \frac{(1+e)^\frac{3}{2}}{(1-e)^\frac{1}{2}} \\
	c_e &= \sqrt{\frac{2}{\pi z}} \left( \frac{1+e}{1-e} \right) ^\frac{1}{2} (1-e^2)
	\end{align}
\end{subequations}
the rate of change for highly eccentric orbits is
\begin{subequations}
	\label{eq:contHighLow}
	\begin{align}
	F_a = \dv{a}{t} &= -k_a c_a [\pmb{r}^T \pmb{K}_a^h \pmb{e} + \mathcal{O}\left(\frac{1}{z^6}\right)] \\
	F_e = \dv{e}{t} &= -k_e c_e [\pmb{r}^T \pmb{K}_e^h \pmb{e} + \mathcal{O}\left(\frac{1}{z^6}\right)]
	\end{align}
\end{subequations}

\section{Jacobian of Dynamics in $a$ and $e$}
\label{sec:jacobian}
The partial derivatives of the dynamics with respect to $a$ and $e$ are given here for the three different regimes discussed in \ref{sec:KH}. The partial derivatives of a partial atmosphere defined in Equation~\ref{eq:smoothAtmosphere} (dropping the subscript $p$), and given $h_p = a (1-e) - R$, can be found as
\begin{subequations}
	\begin{align}
	\pdv{\rho(h_p)}{a} &= -\frac{1-e}{H} \rho(h_p) \\
	\pdv{\rho(h_p)}{e} &= \frac{a}{H} \rho(h_p)
	\end{align}
\end{subequations}
Thus, the partial derivatives of $k_a$ and $k_e$ (see Equation~\ref{eq:terms}) with respect to $a$ and $e$ are
\begin{subequations}
	\label{eq:termder}
	\begin{align}
	\pdv{k_a}{a} &= \delta \sqrt{\mu} \left( \frac{\rho(h_p)}{2 \sqrt{a}} + \sqrt{a} \pdv{\rho(h_p)}{a} \right) = k_a \left(\frac{1}{2 a} - \frac{1-e}{H} \right) \\
	\pdv{k_a}{e} &= \delta \sqrt{\mu a} \pdv{\rho(h_p)}{e} = k_a \frac{a}{H} \\
	\pdv{k_e}{a} &= \delta \sqrt{\mu} \left(-\frac{\rho(h_p)}{2 a^{\frac{3}{2}}} + \frac{1}{\sqrt{a}} \pdv{\rho(h_p)}{a} \right) = k_e \left( -\frac{1}{2a}-\frac{1-e}{H} \right) \\
	\pdv{k_e}{e} &= \delta \sqrt{\frac{\mu}{a}} \pdv{\rho(h_p)}{e} = k_e \frac{a}{H}
	\end{align}
\end{subequations}

\subsection{Circular Orbit}
For circular orbits, the rate and derivative in $e$ vanishes and the partial derivative of $F_a$ with respect to $a$, combining Equations~\ref{eq:contCircDer} and~\ref{eq:termder}, is
\begin{equation}
	\pdv{F_a}{a} = \left( \frac{1}{2a} - \frac{1}{H} \right) F_a
\end{equation}

\subsection{Low Eccentric Orbit}
For low eccentric orbits, with $e \leq e_b$, the partial derivative of $F_a$ and $F_e$ with respect to $a$ and $e$, combining Equations~\ref{eq:contLowDer} and~\ref{eq:termder} and using the product rule, are
\begin{subequations}
	\begin{align}
		\pdv{F_a}{a} 
		&= \left(\frac{1}{2a} - \frac{1}{H} \right) F_a - k_a \exp(-z) \pmb{e}^T \pmb{K}_a^l \frac{e}{H} \pdv{\pmb{I}}{z} \\
		\pdv{F_a}{e} 
		&= -k_a \exp(-z) \left[ \pdv{\pmb{e}^T}{e} \pmb{K}_a^l \pmb{I} + \pmb{e}^T \pmb{K}_a^l \frac{a}{H} \pdv{\pmb{I}}{z} \right] \\
		\pdv{F_e}{a} 
		&= \left(-\frac{1}{2a} - \frac{1}{H} \right) F_e - k_e \exp(-z) \pmb{e}^T \pmb{K}_e^l \frac{e}{H} \pdv{\pmb{I}}{z}  \\
		\pdv{F_e}{e} 
		&= -k_e \exp(-z) \left[ \pdv{\pmb{e}^T}{e} \pmb{K}_e^l \pmb{I} +  \pmb{e}^T \pmb{K}_e^l \frac{a}{H} \pdv{\pmb{I}}{z} \right]
	\end{align}
\end{subequations}
where
\begin{equation}
	\pdv{I_n(z)}{z} = \frac{1}{2} \left( I_{n-1}(z) + I_{n+1}(z) \right) \qquad \pdv{I_0(z)}{z} = I_1(z)
\end{equation}
and
\begin{equation}
	\pdv{(e^n)}{e} = n e^{n-1}
\end{equation}

\subsection{High Eccentric Orbit}
Using the partial derivatives of $c_a$ and $c_e$ from Equation~\ref{eq:termsHigh} with respect to $a$ and $e$
\begin{subequations}
	\begin{align}
	\pdv{c_a}{a} &= c_a \left( -\frac{1}{2a} \right) \\
	\pdv{c_a}{e} &= c_a \left( -\frac{1-4e+e^2}{2e(1-e^2)} \right) \\
	\pdv{c_e}{a} &= c_e \left( -\frac{1}{2a} \right) \\
	\pdv{c_e}{e} &= c_e \left( -\frac{1-2e+3e^2}{2e(1-e^2)} \right)
	\end{align}
\end{subequations}
it follows that
\begin{subequations}
	\begin{align}
	\pdv{}{a} \left( k_a  c_a \right)
	&= k_a c_a \left( -\frac{1-e}{H} \right) \\
	\pdv{}{e} \left( k_a  c_a \right)
	&= k_a c_a \left( \frac{a}{H} - \frac{1-4e+e^2}{2e(1-e^2)} \right) \\
	\pdv{}{a} \left( k_e  c_e \right)
	&= k_e c_e \left( -\frac{1}{a} -\frac{1-e}{H} \right) \\
	\pdv{}{e} \left( k_e c_e \right) 
	&= k_e c_e \left(\frac{a}{H}-\frac{1-2e+3e^2}{2e(1-e^2)} \right)
	\end{align}
\end{subequations}
and the partial derivatives of $F_a$ and $F_e$ for high eccentric orbits (see Equation~\ref{eq:contHighLow}), with $e \geq e_b$, with respect to $a$ and $e$ become
\begin{subequations}
	\begin{align}
	\pdv{F_a}{a}
	&= \left( -\frac{1-e}{H} \right) F_a - k_a c_a \pdv{\pmb{r}^T}{a} \pmb{K}_a^h \pmb{e} \\
	\pdv{F_a}{e}
	&= \left( \frac{a}{H} - \frac{1-4e+e^2}{2e(1-e^2)} \right) F_a - k_a c_a \left[ \pdv{\pmb{r}^T}{e} \pmb{K}_a^h \pmb{e} + \pmb{r}^T \pmb{K}_a^h \pdv{\pmb{e}}{e} \right] \\
	\pdv{F_e}{a}
	&= \left( -\frac{1}{a} -\frac{1-e}{H} \right) F_e - k_e c_e \pdv{\pmb{r}^T}{a} \pmb{K}_e^h \pmb{e} \\
	\pdv{F_e}{e}
	&= \left(\frac{a}{H}-\frac{1-2e+3e^2}{2e(1-e^2)} \right) F_e - k_e c_e \left[ \pdv{\pmb{r}^T}{e} \pmb{K}_e^h \pmb{e} + \pmb{r}^T \pmb{K}_e^h \pdv{\pmb{e}}{e} \right]
	\end{align}
\end{subequations}
where
\begin{subequations}
	\begin{align}
	r_n &= l^{-n} = \left( \frac{ae}{H} (1-e^2) \right)^{-n} \\
	\pdv{r_n}{a} &= -n l^{-(n+1)} \frac{e}{H} (1-e^2) = -\frac{n}{a} r_n \\
	\pdv{r_n}{e} &= -n l^{-(n+1)} \frac{a}{H} (1-3e^2) = -\frac{n(1-3e^2)}{e(1-e^2)} r_n
	\end{align}
\end{subequations}

\section{Variable Atmosphere Model Parameters}
\label{sec:variableAtmosphereParameters}
Tables~\ref{tab:fitParametersVariableA} and~\ref{tab:fitParametersVariableB} list the parameters to calculate $\pmb{a}$ and $\pmb{b}$ according to Equation~\ref{eq:abvectors} as a function of the normalised $\tilde{T}_\infty$. The two vectors are needed to recover $\hat{\rho}$${}_p$ and $H$${}_p$ $\forall p$, according to Equation~\ref{eq:varparams}.
Note that the model should only be used for $T_\infty \in [T_0=650, T_1=1350]$~K. 

\begin{table}
	\centering
	\caption{Parameters to calculate $\pmb{a}$ as a function of $\tilde{T}_\infty$. The factors are of unit [km$^{-1}$].}
	\newcommand{\ccell}[1]{\multicolumn{1}{c}{#1}}

\begin{tabular}{l r r r}
	
	\hline
	$p$	& \ccell{$a_{p0}$}	& \ccell{$a_{p1}$}	& \ccell{$a_{p2}$} \\
	\hline
	
	$1$	& $-1.98541\text{e}-1$	& $-1.40701\text{e}-2$	& $ 1.87647\text{e}-2$	\\
	$2$	& $-9.71648\text{e}-2$	& $ 7.16062\text{e}-3$	& $ 4.77822\text{e}-2$	\\
	$3$	& $-5.05069\text{e}-2$	& $ 3.33725\text{e}-2$	& $-1.85987\text{e}-2$	\\
	$4$	& $-2.83356\text{e}-2$	& $ 1.64584\text{e}-2$	& $-3.32683\text{e}-2$	\\
	$5$	& $-2.18893\text{e}-2$	& $ 8.84693\text{e}-3$	& $ 5.46460\text{e}-2$ \\
	$6$	& $-6.24488\text{e}-3$	& $ 4.90041\text{e}-3$	& $-6.03999\text{e}-3$	\\
	$7$	& $-2.82771\text{e}-3$	& $-3.17505\text{e}-3$	& $ 1.93697\text{e}-3$	\\
	$8$	& $-8.53512\text{e}-4$	& $ 7.92640\text{e}-4$	& $-1.24063\text{e}-3$	\\
	
	\hline
	$p$	& \ccell{$a_{p3}$}	& \ccell{$a_{p4}$}	& \ccell{$a_{p5}$} \\
	\hline
	
	$1$	& $-1.72925\text{e}-2$	& $ 2.77798\text{e}-2$	& $-9.95750\text{e}-2$ \\
	$2$	& $-1.51184\text{e}-1$	& $ 3.51432\text{e}-1$	& $-7.02642\text{e}-1$ \\
	$3$	& $-1.03728\text{e}-1$	& $ 5.51289\text{e}-1$	& $-1.41638\text{e}+0$ \\
	$4$	& $ 8.69501\text{e}-2$	& $-6.20406\text{e}-2$	& $-3.36952\text{e}-1$ \\
	$5$	& $-2.34999\text{e}-1$	& $ 5.47095\text{e}-1$	& $-8.27779\text{e}-1$ \\
	$6$	& $-7.24190\text{e}-2$	& $ 5.32824\text{e}-1$	& $-1.79828\text{e}+0$ \\
	$7$	& $ 4.29619\text{e}-2$	& $-1.78919\text{e}-1$	& $ 3.53528\text{e}-1$ \\
	$8$	& $ 4.65874\text{e}-3$	& $-1.87465\text{e}-2$	& $ 8.70408\text{e}-3$ \\
	
	\hline
	$p$	& \ccell{$a_{p6}$}	& \ccell{$a_{p7}$}	& \ccell{$a_{p8}$} \\
	\hline
	
	$1$	& $ 1.76679\text{e}-1$	& $-1.37542\text{e}-1$	& $ 3.94618\text{e}-2$	\\
	$2$	& $ 9.01640\text{e}-1$	& $-6.03103\text{e}-1$	& $ 1.59691\text{e}-1$	\\
	$3$	& $ 1.87770\text{e}+0$	& $-1.22379\text{e}+0$	& $ 3.11852\text{e}-1$	\\
	$4$	& $ 8.28293\text{e}-1$	& $-6.99209\text{e}-1$	& $ 2.06734\text{e}-1$	\\
	$5$	& $ 7.76841\text{e}-1$	& $-4.02671\text{e}-1$	& $ 8.74533\text{e}-2$	\\
	$6$	& $ 2.85818\text{e}+0$	& $-2.11311\text{e}+0$	& $ 5.91400\text{e}-1$	\\
	$7$	& $-3.82857\text{e}-1$	& $ 2.16923\text{e}-1$	& $-5.02721\text{e}-2$	\\
	$8$	& $ 3.62357\text{e}-2$	& $-4.73838\text{e}-2$	& $ 1.66805\text{e}-2$	\\
	
	\hline
	
\end{tabular}
	\label{tab:fitParametersVariableA}
\end{table}

\begin{table}
	\centering
	\caption{Parameters to calculate $\pmb{b}$ as a function of $\tilde{T}_\infty$. The factors are of unit [ln(kg/m$^3$)].}
	\newcommand{\ccell}[1]{\multicolumn{1}{c}{#1}}

\begin{tabular}{l r r r}
	
	\hline
	$p$	& \ccell{$b_{p0}$}	& \ccell{$b_{p1}$}	& \ccell{$b_{p2}$}	\\
	\hline
	
	$1$	& $ 5.35674\text{e}+0$	& $ 1.36142\text{e}+0$	& $-1.71993\text{e}+0$	\\
	$2$	& $-6.96022\text{e}+0$	& $-1.71534\text{e}-1$	& $-6.26282\text{e}+0$	\\
	$3$	& $-1.33334\text{e}+1$	& $-4.29240\text{e}+0$	& $ 1.12545\text{e}+0$	\\
	$4$	& $-1.78792\text{e}+1$	& $-2.89047\text{e}+0$	& $ 3.93500\text{e}+0$	\\
	$5$	& $-2.09320\text{e}+1$	& $ 8.52674\text{e}+0$	& $-5.08863\text{e}+1$	\\
	$6$	& $-2.93700\text{e}+1$	& $ 5.68339\text{e}-2$	& $-2.61029\text{e}+1$	\\
	$7$	& $-3.29807\text{e}+1$	& $ 4.90080\text{e}+0$	& $ 1.78391\text{e}+1$	\\
	$8$	& $-3.51561\text{e}+1$	& $-2.66659\text{e}+0$	& $ 1.73783\text{e}+0$	\\
	
	\hline
	$p$	& \ccell{$b_{p3}$}	& \ccell{$b_{p4}$}	& \ccell{$b_{p5}$}	\\
	\hline
	
	$1$	& $ 1.48408\text{e}+0$	& $-2.43815\text{e}+0$	& $ 9.19988\text{e}+0$	\\
	$2$ & $ 1.70218\text{e}+1$	& $-3.66333\text{e}+1$	& $ 7.26606\text{e}+1$\\
	$3$	& $ 1.41418\text{e}+1$	& $-6.27283\text{e}+1$	& $ 1.53398\text{e}+2$ \\
	$4$	& $ 1.67754\text{e}+1$	& $-1.15289\text{e}+2$	& $ 3.24667\text{e}+2$\\
	$5$	& $ 1.56893\text{e}+2$	& $-3.21951\text{e}+2$	& $ 4.61948\text{e}+2$	\\
	$6$	& $ 2.90804\text{e}+2$	& $-1.47321\text{e}+3$	& $ 3.87334\text{e}+3$ \\
	$7$	& $-9.35850\text{e}+1$	& $ 2.24591\text{e}+2$	& $-3.60868\text{e}+2$ \\
	$8$	& $-4.98942\text{e}+0$	& $ 2.71676\text{e}+1$	& $ 4.15537\text{e}+1$\\
	
	\hline
	$p$	& \ccell{$b_{p6}$}	& \ccell{$b_{p7}$}	& \ccell{$b_{p8}$} \\
	\hline
	
	$1$	& $-1.64492\text{e}+1$	& $ 1.28147\text{e}+1$	& $-3.67526\text{e}+0$	\\
	$2$	& $-9.47544\text{e}+1$	& $ 6.43396\text{e}+1$	& $-1.72245\text{e}+1$	\\
	$3$	& $-2.00134\text{e}+2$	& $ 1.29740\text{e}+2$	& $-3.30267\text{e}+1$	\\
	$4$	& $-4.59063\text{e}+2$	& $ 3.15704\text{e}+2$	& $-8.42405\text{e}+1$	\\
	$5$	& $-4.34126\text{e}+2$	& $ 2.32404\text{e}+2$	& $-5.27733\text{e}+1$	\\
	$6$	& $-5.21125\text{e}+3$	& $ 3.43718\text{e}+3$	& $-8.85649\text{e}+2$	\\
	$7$	& $ 3.73065\text{e}+2$	& $-2.15221\text{e}+2$	& $ 5.18052\text{e}+1$	\\
	$8$	& $-1.88208\text{e}+2$	& $ 1.86631\text{e}+2$	& $-5.96266\text{e}+1$	\\
	
	\hline
	
\end{tabular}
	\label{tab:fitParametersVariableB}
\end{table}

\end{document}